\documentclass[12pt,twocolumn]{aastex631}
\usepackage{hyperref}
\usepackage{color}
\usepackage{natbib}
\usepackage{graphicx}
\usepackage{xfrac}
\usepackage{ulem}
\usepackage{xcolor}
\usepackage{amsmath}
\usepackage{longtable}

\shorttitle{DESI Quasar Visual Inspection}
\shortauthors{Alexander et~al.}

\begin{document}
\title{The DESI Survey Validation: Results from  Visual Inspection of the Quasar Survey Spectra}
\correspondingauthor{David~M.~Alexander}
\email{d.m.alexander@durham.ac.uk}

\author[0000-0002-5896-6313]{David~M.~Alexander}
\affiliation{Centre for Extragalactic Astronomy, Department of Physics, Durham University, South Road, Durham, DH1 3LE, UK}
\author[0000-0002-4213-8783]{Tamara~M.~Davis}
\affiliation{School of Mathematics and Physics, University of Queensland, 4072, Australia}
\author[0000-0001-8996-4874]{E.~Chaussidon}
\affiliation{IRFU, CEA, Universit\'{e} Paris-Saclay, F-91191 Gif-sur-Yvette, France}
\author[0000-0003-1251-532X]{V.~A.~Fawcett}
\affiliation{Centre for Extragalactic Astronomy, Department of Physics, Durham University, South Road, Durham, DH1 3LE, UK}
\author[0000-0003-4089-6924]{Alma  X.~Gonzalez-Morales}
\affiliation{Consejo Nacional de Ciencia y Tecnolog\'{\i}a, Av. Insurgentes Sur 1582. Colonia Cr\'{e}dito Constructor, Del. Benito Ju\'{a}rez C.P. 03940, M\'{e}xico D.F. M\'{e}xico}
\affiliation{Departamento de F\'{i}sica, Universidad de Guanajuato - DCI, C.P. 37150, Leon, Guanajuato, M\'{e}xico}
\author[0000-0001-8857-7020]{Ting-Wen Lan}
\affiliation{Graduate Institute of Astrophysics and Department of Physics, National Taiwan University, No. 1, Sec. 4, Roosevelt Rd., Taipei 10617, Taiwan}
\author{Christophe~Yèche}
\affiliation{IRFU, CEA, Universit\'{e} Paris-Saclay, F-91191 Gif-sur-Yvette, France}
\author[0000-0001-6098-7247]{S.~Ahlen}
\affiliation{Physics Dept., Boston University, 590 Commonwealth Avenue, Boston, MA 02215, USA}
\author{J.~N.~Aguilar}
\affiliation{Lawrence Berkeley National Laboratory, 1 Cyclotron Road, Berkeley, CA 94720, USA}
\author{E.~Armengaud}
\affiliation{IRFU, CEA, Universit\'{e} Paris-Saclay, F-91191 Gif-sur-Yvette, France}
\author[0000-0003-4162-6619]{S.~Bailey}
\affiliation{Lawrence Berkeley National Laboratory, 1 Cyclotron Road, Berkeley, CA 94720, USA}
\author{D.~Brooks}
\affiliation{Department of Physics \& Astronomy, University College London, Gower Street, London, WC1E 6BT, UK}
\author{Z.~Cai}
\affiliation{Department of Astronomy and Astrophysics, University of California, Santa Cruz, 1156 High Street, Santa Cruz, CA 95065, USA}
\affiliation{Department of Astronomy, Tsinghua University, 30 Shuangqing Road, Haidian District, Beijing, China, 100190}
\author{R.~Canning}
\affiliation{Institute of Cosmology \& Gravitation, University of Portsmouth, Dennis Sciama Building, Portsmouth, PO1 3FX, UK}
\author[0000-0003-4074-5659]{A.~Carr}
\affiliation{School of Mathematics and Physics, University of Queensland, 4072, Australia}
\author[0000-0002-5692-5243]{S.~Chabanier}
\affiliation{Lawrence Berkeley National Laboratory, 1 Cyclotron Road, Berkeley, CA 94720, USA}
\author{Marie-Claude Cousinou}
\affiliation{Aix Marseille Univ, CNRS/IN2P3, CPPM, Marseille, France}
\author{K.~Dawson}
\affiliation{Department of Physics and Astronomy, The University of Utah, 115 South 1400 East, Salt Lake City, UT 84112, USA}
\author{A.~de la Macorra}
\affiliation{Instituto de F\'{\i}sica, Universidad Nacional Aut\'{o}noma de M\'{e}xico,  Cd. de M\'{e}xico  C.P. 04510,  M\'{e}xico}
\author{A.~Dey}
\affiliation{NSF's National Optical-Infrared Astronomy Research Laboratory, 950 N. Cherry Avenue, Tucson, AZ 85719, USA}
\author[0000-0002-5665-7912]{Biprateep~Dey}
\affiliation{Department of Physics \& Astronomy and Pittsburgh Particle Physics, Astrophysics, and Cosmology Center (PITT PACC), University of Pittsburgh, 3941 O'Hara Street, Pittsburgh, PA 15260, USA}
\author[0000-0002-5402-1216]{G.~Dhungana}
\affiliation{Department of Physics, Southern Methodist University, 3215 Daniel Avenue, Dallas, TX 75275, USA}
\author{A.~C.~Edge}
\affiliation{Centre for Extragalactic Astronomy, Department of Physics, Durham University, South Road, Durham, DH1 3LE, UK}
\author{S.~Eftekharzadeh}
\affiliation{Universities Space Research Association, NASA Ames Research Centre}
\author{K.~Fanning}
\affiliation{Department of Physics, University of Michigan, Ann Arbor, MI 48109, USA}
\affiliation{Physics Department, University of Michigan Ann Arbor, MI 48109, USA}
\author{James Farr}
\affiliation{Department of Physics \& Astronomy, University College London, Gower Street, London, WC1E 6BT, UK}
\author{A.~Font-Ribera}
\affiliation{Institut de F\'{i}sica d’Altes Energies (IFAE), The Barcelona Institute of Science and Technology, Campus UAB, 08193 Bellaterra Barcelona, Spain}
\author{J.~Garcia-Bellido}
\affiliation{Instituto de F\'{\i}sica Te\'{o}rica (IFT) UAM/CSIC, Universidad Aut\'{o}noma de Madrid, Cantoblanco, E-28049, Madrid, Spain}
\author{Lehman Garrison}
\affiliation{Center for Computational Astrophysics Flatiron Institute 162 5th Ave., New York, NY 10010, USA}
\author{E.~Gaztañaga}
\affiliation{Institute of Space Sciences, ICE-CSIC, Campus UAB, Carrer de Can Magrans s/n, 08913 Bellaterra, Barcelona, Spain}
\author[0000-0003-3142-233X]{Satya~Gontcho A Gontcho}
\affil{Lawrence Berkeley National Laboratory, One Cyclotron Road, Berkeley, CA 94720, USA}
\affil{Department of Physics and Astronomy, University of Rochester, 500 Joseph C. Wilson Boulevard, Rochester, NY 14627, USA}
\author{C.~Gordon}
\affiliation{Institut de F\'{i}sica d’Altes Energies (IFAE), The Barcelona Institute of Science and Technology, Campus UAB, 08193 Bellaterra Barcelona, Spain}
\author{Stefany Guadalupe Medellin Gonzalez}
\affiliation{Departamento de F\'{i}sica, Universidad de Guanajuato - DCI, C.P. 37150, Leon, Guanajuato, M\'{e}xico}
\author{J.~Guy}
\affiliation{Lawrence Berkeley National Laboratory, 1 Cyclotron Road, Berkeley, CA 94720, USA}
\author[0000-0002-9136-9609]{Hiram K. Herrera-Alcantar}
\affiliation{Departamento de F\'{i}sica, Universidad de Guanajuato - DCI, C.P. 37150, Leon, Guanajuato, M\'{e}xico}
\author[0000-0003-4176-6486]{L.~Jiang}
\affiliation{Kavli Institute for Astronomy and Astrophysics at Peking University, PKU, 5 Yiheyuan Road, Haidian District, Beijing 100871, P.R. China}
\author{S.~Juneau}
\affiliation{NSF's National Optical-Infrared Astronomy Research Laboratory, 950 N. Cherry Avenue, Tucson, AZ 85719, USA}
\author{N.G.~Kara{\c c}ayl{\i}}
\affiliation{Department of Physics, The Ohio State University, 191 West Woodruff Avenue, Columbus, OH 43210, USA}
\affiliation{Center for Cosmology and AstroParticle Physics, The Ohio State University, 191 West Woodruff Avenue, Columbus, OH 43210, USA}
\affiliation{Department of Astronomy, The Ohio State University, 4055 McPherson Laboratory, 140 W 18th Avenue, Columbus, OH 43210, USA}
\author{R.~Kehoe}
\affiliation{Department of Physics, Southern Methodist University, 3215 Daniel Avenue, Dallas, TX 75275, USA}
\author[0000-0003-3510-7134]{T.~Kisner}
\affiliation{Lawrence Berkeley National Laboratory, 1 Cyclotron Road, Berkeley, CA 94720, USA}
\author[0000-0002-5825-579X]{A.~Kov\'acs}
\affiliation{Departamento de Astrof\'{\i}sica, Universidad de La Laguna (ULL), E-38206, La Laguna, Tenerife, Spain}
\affiliation{Instituto de Astrof\'{i}sica de Canarias, C/ Vía L\'{a}ctea, s/n, 38205 San Crist\'{o}bal de La Laguna, Santa Cruz de Tenerife, Spain}
\author[0000-0003-1838-8528]{M.~Landriau}
\affiliation{Lawrence Berkeley National Laboratory, 1 Cyclotron Road, Berkeley, CA 94720, USA}
\author[0000-0003-1887-1018]{Michael E.~Levi}
\affiliation{Lawrence Berkeley National Laboratory, 1 Cyclotron Road, Berkeley, CA 94720, USA}
\author{C.~Magneville}
\affiliation{IRFU, CEA, Universit\'{e} Paris-Saclay, F-91191 Gif-sur-Yvette, France}
\author[0000-0002-4279-4182]{P.~Martini}
\affiliation{Center for Cosmology and AstroParticle Physics, The Ohio State University, 191 West Woodruff Avenue, Columbus, OH 43210, USA}
\affiliation{Department of Astronomy, The Ohio State University, 4055 McPherson Laboratory, 140 W 18th Avenue, Columbus, OH 43210, USA}
\author[0000-0002-1125-7384]{Aaron M. Meisner}
\affiliation{NSF's National Optical-Infrared Astronomy Research Laboratory, 950 N. Cherry Avenue, Tucson, AZ 85719, USA}
\author{M.~Mezcua}
\affiliation{Institute of Space Sciences, ICE-CSIC, Campus UAB, Carrer de Can Magrans s/n, 08913 Bellaterra, Barcelona, Spain}
\author{R.~Miquel}
\affiliation{Instituci\'{o} Catalana de Recerca i Estudis Avan\c{c}ats, Passeig de Llu\'{\i}s Companys, 23, 08010 Barcelona, Spain}
\affiliation{Institut de F\'{i}sica d’Altes Energies (IFAE), The Barcelona Institute of Science and Technology, Campus UAB, 08193 Bellaterra Barcelona, Spain}
\author[0000-0002-6998-6678]{P.~Montero Camacho}
\affiliation{Department of Astronomy, Tsinghua University, 30 Shuangqing Road, Haidian District, Beijing, China, 100190}
\author[0000-0002-2733-4559]{J.~Moustakas}
\affiliation{Department of Physics and Astronomy, Siena College, 515 Loudon Road, Loudonville, NY 12211, USA}
\author{Andrea Muñoz-Gutiérrez}
\affiliation{Instituto de F\'{\i}sica, Universidad Nacional Aut\'{o}noma de M\'{e}xico,  Cd. de M\'{e}xico  C.P. 04510,  M\'{e}xico}
\author{Adam~D.~Myers}
\affiliation{Department of Physics \& Astronomy, University  of Wyoming, 1000 E. University, Dept.~3905, Laramie, WY 82071, USA}
\author{S.~Nadathur}
\affiliation{Department of Physics \& Astronomy, University College London, Gower Street, London, WC1E 6BT, UK}
\affiliation{Institute of Cosmology \& Gravitation, University of Portsmouth, Dennis Sciama Building, Portsmouth, PO1 3FX, UK}
\author{L.~Napolitano}
\affiliation{Department of Physics \& Astronomy, University  of Wyoming, 1000 E. University, Dept.~3905, Laramie, WY 82071, USA}
\author[0000-0001-6590-8122]{J.~D.~Nie}
\affiliation{National Astronomical Observatories, Chinese Academy of Sciences, A20 Datun Rd., Chaoyang District, Beijing, 100012, P.R. China}
\author[0000-0003-3188-784X]{N.~Palanque-Delabrouille}
\affiliation{IRFU, CEA, Universit\'{e} Paris-Saclay, F-91191 Gif-sur-Yvette, France}
\affiliation{Lawrence Berkeley National Laboratory, 1 Cyclotron Road, Berkeley, CA 94720, USA}
\author{Z.~Pan}
\affiliation{Kavli Institute for Astronomy and Astrophysics at Peking University, PKU, 5 Yiheyuan Road, Haidian District, Beijing 100871, P.R. China}
\author[0000-0002-0644-5727]{W.J.~Percival}
\affiliation{Department of Physics and Astronomy, University of Waterloo, 200 University Ave W, Waterloo, ON N2L 3G1, Canada}
\affiliation{Perimeter Institute for Theoretical Physics, 31 Caroline St. North, Waterloo, ON N2L 2Y5, Canada}
\affiliation{Waterloo Centre for Astrophysics, University of Waterloo, 200 University Ave W, Waterloo, ON N2L 3G1, Canada}
\author[0000-0001-6979-0125]{I.~P\'erez-R\`afols}
\affiliation{Institut de F\'{i}sica d’Altes Energies (IFAE), The Barcelona Institute of Science and Technology, Campus UAB, 08193 Bellaterra Barcelona, Spain}
\affiliation{Sorbonne Universit\'{e}, CNRS/IN2P3, Laboratoire de Physique Nucl\'{e}aire et de Hautes Energies (LPNHE), FR-75005 Paris, France}
\author{C.~Poppett}
\affiliation{Lawrence Berkeley National Laboratory, 1 Cyclotron Road, Berkeley, CA 94720, USA}
\affiliation{Space Sciences Laboratory, University of California, Berkeley, 7 Gauss Way, Berkeley, CA  94720, USA}
\affiliation{University of California, Berkeley, 110 Sproul Hall \#5800 Berkeley, CA 94720, USA}
\author[0000-0001-7145-8674]{F.~Prada}
\affiliation{Instituto de Astrofisica de Andaluc\'{i}a, Glorieta de la Astronom\'{i}a, s/n, E-18008 Granada, Spain}
\author{C\'esar Ram\'irez-P\'erez}
\affiliation{Institut de F\'{i}sica d’Altes Energies (IFAE), The Barcelona Institute of Science and Technology, Campus UAB, 08193 Bellaterra Barcelona, Spain}
\author{C.~Ravoux}
\affiliation{IRFU, CEA, Universit\'{e} Paris-Saclay, F-91191 Gif-sur-Yvette, France}
\author{D.~J.~Rosario}
\affiliation{Centre for Extragalactic Astronomy, Department of Physics, Durham University, South Road, Durham, DH1 3LE, UK}
\author{M.~Schubnell}
\affiliation{Department of Physics, University of Michigan, Ann Arbor, MI 48109, USA}
\affiliation{Physics Department, University of Michigan Ann Arbor, MI 48109, USA}
\author[0000-0003-1704-0781]{Gregory~Tarl\'{e}}
\affiliation{Department of Physics, University of Michigan, Ann Arbor, MI 48109, USA}
\author[0000-0002-1748-3745]{M.~Walther}
\affiliation{Excellence Cluster ORIGINS, Boltzmannstrasse 2, D-85748 Garching, Germany}
\affiliation{University Observatory, Faculty of Physics, Ludwig-Maximilians-Universit\"{a}t, Scheinerstr. 1, 81677 M\"{u}nchen, Germany}
\author{B.~Weiner}
\affiliation{Steward Observatory, University of Arizona, 933 N, Cherry Ave, Tucson, AZ 85721, USA}
\author[0000-0002-7520-5911]{S.~Youles}
\affiliation{Institute of Cosmology \& Gravitation, University of Portsmouth, Dennis Sciama Building, Portsmouth, PO1 3FX, UK}
\author{Zhimin~Zhou}
\affiliation{National Astronomical Observatories, Chinese Academy of Sciences, A20 Datun Rd., Chaoyang District, Beijing, 100012, P.R. China}
\author[0000-0002-6684-3997]{H.~Zou}
\affiliation{National Astronomical Observatories, Chinese Academy of Sciences, A20 Datun Rd., Chaoyang District, Beijing, 100012, P.R. China}
\author{Siwei Zou}
\affiliation{Department of Astronomy, Tsinghua University, 30 Shuangqing Road, Haidian District, Beijing, China, 100190}
\affiliation{Kavli Institute for Astronomy and Astrophysics at Peking University, PKU, 5 Yiheyuan Road, Haidian District, Beijing 100871, P.R. China}

\begin{abstract}
A key component of the Dark Energy Spectroscopic Instrument (DESI) survey validation (SV) is a detailed visual inspection (VI) of the optical spectroscopic data to quantify key survey metrics. In this paper we present results from VI of the quasar survey using deep coadded SV spectra. We show that the majority ($\approx$~70\%) of the main-survey targets are spectroscopically confirmed as quasars, with $\approx$~16\% galaxies, $\approx$~6\% stars, and $\approx$~8\% low-quality spectra lacking reliable features. A non-negligible fraction of the quasars are misidentified by the standard spectroscopic pipeline but we show that the majority can be recovered using post-pipeline ``afterburner" quasar-identification approaches. We combine these ``afterburners" with our standard pipeline to create a modified pipeline to increase the overall quasar yield. At the depth of the main DESI survey both pipelines achieve a good-redshift purity (reliable redshifts measured within 3000~km~s$^{-1}$) of $\approx$~99\%; however, the modified pipeline recovers $\approx$~94\% of the visually inspected quasars, as compared to $\approx$~86\% from the standard pipeline. We demonstrate that both pipelines achieve a median redshift precision and accuracy of $\approx$~100~km~s$^{-1}$ and $\approx$~70~km~s$^{-1}$, respectively. We constructed composite spectra to investigate why some quasars are missed by the standard pipeline and find that they are more host-galaxy dominated (i.e.,\ distant analogs of ``Seyfert galaxies") and/or more dust reddened than the standard-pipeline quasars. We also show example spectra to demonstrate the overall diversity of the DESI quasar sample and provide strong-lensing candidates where two targets contribute to a single spectrum.
\end{abstract}


\section{Introduction}

DESI (the Dark Energy Spectroscopic Instrument; Levi et~al.\ 2013) is a Stage IV dark-energy experiment. It is the successor to several Stage III experiments, including the BOSS/eBOSS spectroscopic redshift surveys (extended Baryonic Oscillation Spectroscopic Survey; Dawson et~al.\ 2013, 2016) and the DES imaging survey (Dark Energy Survey; Dark Energy Survey Collaboration et~al.\ 2016a,b). DESI is also complementary to the Vera C. Rubin Observatory, an up-coming Stage IV imaging experiment (Ivezic et~al.\ 2019), {\it Euclid} (Euclid Collaboration et~al.\ 2020), and future large spectroscopic surveys (e.g.,\ 4MOST and WEAVE; Pieri et~al.\ 2016, Smith et~al.\ 2016, Driver et~al.\ 2019, Merloni et~al.\ 2019, Richard et~al.\ 2019). 

The primary objective of DESI is to study Baryon Acoustic Oscillations (BAO) and to measure the growth of structure through redshift-space distortions (RSD), although the extensive DESI dataset permits many additional ground-breaking cosmological and astrophysical experiments (see DESI collaboration et~al.\ 2016a). To achieve these aims DESI is undertaking a large-area ($\approx$~10,000--14,000~deg$^2$) and sensitive ($r\approx$~23~mag) spectroscopic survey of $\approx$~40~million galaxies and $\approx$~3~million quasars over the next 5~years.

DESI is the superposition of four individual spectroscopic surveys to probe a wide range in redshift and mass: the Bright Galaxy Survey (BGS; Hahn et~al.\ 2022), the Luminous Red Galaxy (LRG; Zhou et~al.\ 2022) survey, the Emission-Line Galaxy survey (ELG; Raichoor et~al.\ 2022a), and the quasar survey (QSO; Chaussidon et~al.\ 2022). The BGS selects the low-$z$ galaxy tail out to $z\approx$~0.5 using a bright optical magnitude-limited target selection, while the LRG and ELG surveys use a combination of optical colour and magnitude thresholds to select massive passive galaxies and star-forming galaxies over $z\approx$~0.4--1.4 and $z\approx$~0.6--1.6, respectively. The QSO survey uses a broad colour-magnitude selection of quasars to trace the widest redshift range (over $z\approx$~0--5) and uniquely probes the high end of the overall redshift distribution in DESI ($z>1.6$). In addition, DESI is targeting $\approx$~7 million stars to enable a comprehensive census and analysis of the Milky Way (Cooper et~al.\ 2022).

The purpose of this paper is to present to the community the results of the visual inspection (VI) of the optical spectra from the QSO survey. VI is a key element of large-scale surveys, even for massive spectroscopic surveys like DESI with 10's of millions of spectra. VI provides a critical evaluation of the performance and development of the pipeline processes, serves as a test bed for the target-selection approaches, and quantifies key survey metrics (e.g.,\ spectroscopic quality and redshift reliability). The majority of the VI in DESI has been focused on the 6~month survey validation (SV) phase where various target-selection approaches are tested and refined in order to achieve the key scientific objectives of DESI (see DESI collaboration et~al. 2022 for a description and overview of the DESI SV). The QSO target selection during SV (referred to as SV1) was more liberal than the target selection in the main DESI survey (referred to as main), probing to fainter optical magnitudes and utilising a looser QSO-selection approach to maximise the diversity and density of targets (see Chaussidon et~al. 2022 for the QSO target selection). The optical spectroscopy during the SV phase was also substantially deeper than that employed in the main DESI survey (exposures up-to 10 times longer), allowing for the construction of accurate and reliable VI ``truth tables" with which to evaluate the overall performance of DESI. 

\begin{table*}
%\begin{center} \
%\begin{longtable*}{c|c|c|c|l}
\caption{Summary of the QSO-survey visual-inspection phases}
\label{table:VI_phases}
%\centerline{
%\begin{tabular}{c|c|c|c|l}
\hskip-2.0cm\begin{tabular}{c|c|cc|l|l}
\hline
VI phase & Date & $N_{\rm tiles}$ & $N_{\rm targets}$ & Description & Data assembly\\
\hline\hline
First round & 04/20 & 1 & 917 & VI of tile 68002 & SV0  \\
Second round & 08/20 & 2 & 1432 & VI of tile 68001 and re-inspection of tile 68002 target subsets & Andes  \\
Deep-field VI & 01/21 & 3 & 3671 & VI of all targets in the SV deep-field tiles & Blanc   \\
 & 04/22 & 3 & 18 & VI of deep-field targets with initially bad fibers & Fuji   \\
Sparse VI & 04/21 & 27 & 2391 & VI of selected target subsets in the SV tiles & Cascades \\
\hline
\end{tabular}
\tablecomments{The columns show a name to identify the QSO-survey VI phase, the approximate date of the observations used in the VI, the number of DESI tiles visually inspected, the total number of targets visually inspected, a brief description of the VI phase, and the name of the internal data assembly associated with the production of the optical spectra used in the VI (see Footnote~5). A tile is defined as a unique DESI field of view.}
%}
%\end{longtable*}
\end{table*}

In this paper we quantify the spectroscopic performance of DESI using both the SV1 and main QSO target-selection approaches at both the full SV depth and the shallower main-survey depth ($\approx$~1000~s). This paper is a companion to both the QSO target selection (Chaussidon et~al.\ 2022) and the galaxy VI (Lan et~al.\ 2022) papers. It provides the empirical validation of the QSO target-selection approach and the motivation for a modified pipeline to optimise the selection and overall QSO yield in DESI. The basic VI approach and calculation of the performance metrics are the same for both the galaxy and QSO surveys. However, due to the large diversity of spectral types found in the QSO survey, a greater emphasis is placed on the optical spectroscopic classification in this paper, and the development and testing of a modified pipeline for the QSO survey (Chaussidon et~al.\ 2022). 

In \S2 we present the various datasets used in the QSO VI and provide a detailed description of the VI approach adopted in DESI. In \S3 we present the basic VI results, focusing on two distinct VI datasets to evaluate the overall performance of the QSO survey using both the SV1 and main target selections. In \S4 we exploit the deep VI data to calculate the performance of the QSO survey at the shallower depth of the main 5~year survey. In \S5 we provide a closer look at the spectroscopic data and focus on the broad  diversity of the selected targets and highlight interesting sources, including examples of two objects contributing to the same optical spectrum. We finally summarize our results in \S6. All quoted magnitudes are given in the AB system.

\section{Visual inspection: data and approach}

In this section we provide an overview of the DESI data and the VI approach adopted in DESI. In \S2.1 we provide a brief summary of the DESI instrument, the QSO target selection, and the standard \emph{Redrock} spectroscopic pipeline, along with the details of the QSO target selection. In \S2.2 we describe the approach and tools used in the VI, and in \S2.3 we summarize the main phases of VI in the QSO survey.

%, focusing on the deep-field VI (\S2.3.1) and sparse VI (\S2.3.2) target selections.

\subsection{DESI data and Redrock spectroscopic identification pipeline}

To assist in the interpretation of the results presented in this paper we provide a brief overview of the DESI instrument, the QSO target selection, and the \emph{Redrock} spectral template--redshift fitting code. We refer the interested reader to the referenced papers for more details.

DESI (DESI collaboration et~al.\ 2016b; Abareshi et~al.\ 2022) is a multi-object spectrograph on the NOAO 4~m Mayall telescope at Kitt Peak in Arizona. It uses 5000 robotically controlled positioners to place fibers across a 7.5~deg$^2$ field of view (Abareshi et~al.\ 2022;  Miller et~al. 2022; Silber et~al. 2022). The optical signal from the fibers is fed to optical spectrographs with sensitivity over 360--980~nm. Each optical spectrograph provides medium-resolution spectra across three channels: $R>2100$ in blue (360--590~nm), $R>3200$ in green (566--722~nm), and $R>4100$ in red (747--980~nm).

The overall scale of the DESI experiment requires various supporting data products and software pipelines including the DESI Legacy Imaging Surveys (Zou et~al. 2017; Dey et~al. 2019; Schlegel et~al. 2022), extensive spectroscopic and template-fitting pipelines (Bailey et~al. 2022; Guy et~al. 2022), and pipelines to assign fibers to targets (Raichoor et~al. 2022b), optimize the tiling and planning of the survey observations (Schlafly et~al. 2022), and select targets for spectroscopic observations (Myers et~al. 2022). 

The standard template-fitting code utilized within DESI is called \emph{Redrock} (Bailey et~al.\ 2022). \emph{Redrock} uses a set of templates to represent the spectral properties of the broad object classes identified in DESI: QSOs, galaxies, and stars.\footnote{The spectral templates utilized by \emph{Redrock} were constructed using spectra from the SDSS collaboration.} Combinations of these spectral templates were constructed to provide composite solutions. \emph{Redrock} determines the best-fitting redshift and template solutions to each DESI spectrum across the full range of redshift--template parameter space, selecting the best-fitting solutions on the basis of the lowest reduced $\chi^2$ values. 

In this paper we refer to both a standard \emph{Redrock} pipeline and a modified pipeline. The standard \emph{Redrock} pipeline is the same as that adopted for the DESI galaxy survey components. The modified pipeline (also referred to as ``QSO maker") also adopts \emph{Redrock}, as in the standard pipeline, but additionally utilizes two QSO ``afterburner" algorithms (QuasarNet and Mg~II afterburner; see \S2.3.2). These ``afterburner" algorithms search for significant QSO features missed by \emph{Redrock} which can consequently result in either (1) the reclassification of a galaxy to a QSO or (2) a revised redshift estimate for an identified QSO. Any revised redshifts from the modified pipeline are then refined using \emph{Redrock} but with a tight prior ($dz=$~0.05 width top-hat function) to prioritize solutions that match the identified features (Chaussidon et~al.\ 2022); see \S3.3 for a more detailed description and calculation of the performance of the modified pipeline with respect to the standard \emph{Redrock} pipeline.

The QSO survey targets are selected using a combination of optical ($g$, $r$, $z$) and mid-IR ($W1$, $W2$) colors to reduce the contamination from stars, which can have similar optical colors to QSOs. To reduce the contamination from galaxies an optical morphology cut is also applied, selecting only sources with a ``PSF" morphology in the DR9 legacy imaging (Dey et~al.\ 2019). To improve the efficiency of the QSO selection a random-forest machine-learning approach is adopted. The random forest is trained on both QSOs and stars, aiming to optimise across 11 parameters (10 colour selections from the optical--mid-IR bands plus r-band magnitude) to preferentially select the former and eliminate the latter. For more details of the QSO target selection and survey strategy see Chaussidon et~al.\ (2022). 

For the visual-inspection data explored in this paper, the targets were selected using the more liberal SV1 selection down to $r\approx$~23.5~mag. However, to replicate the expected results in the main DESI survey, we also applied the more conservative main-survey target selection down to $r\approx$~23~mag, which we refer to here as the ``main target selection" but is occasionally referred to elsewhere as SV3 (Chaussidon et~al.\ 2022).

\begin{figure*}
\center
\includegraphics[width=1.0\textwidth]{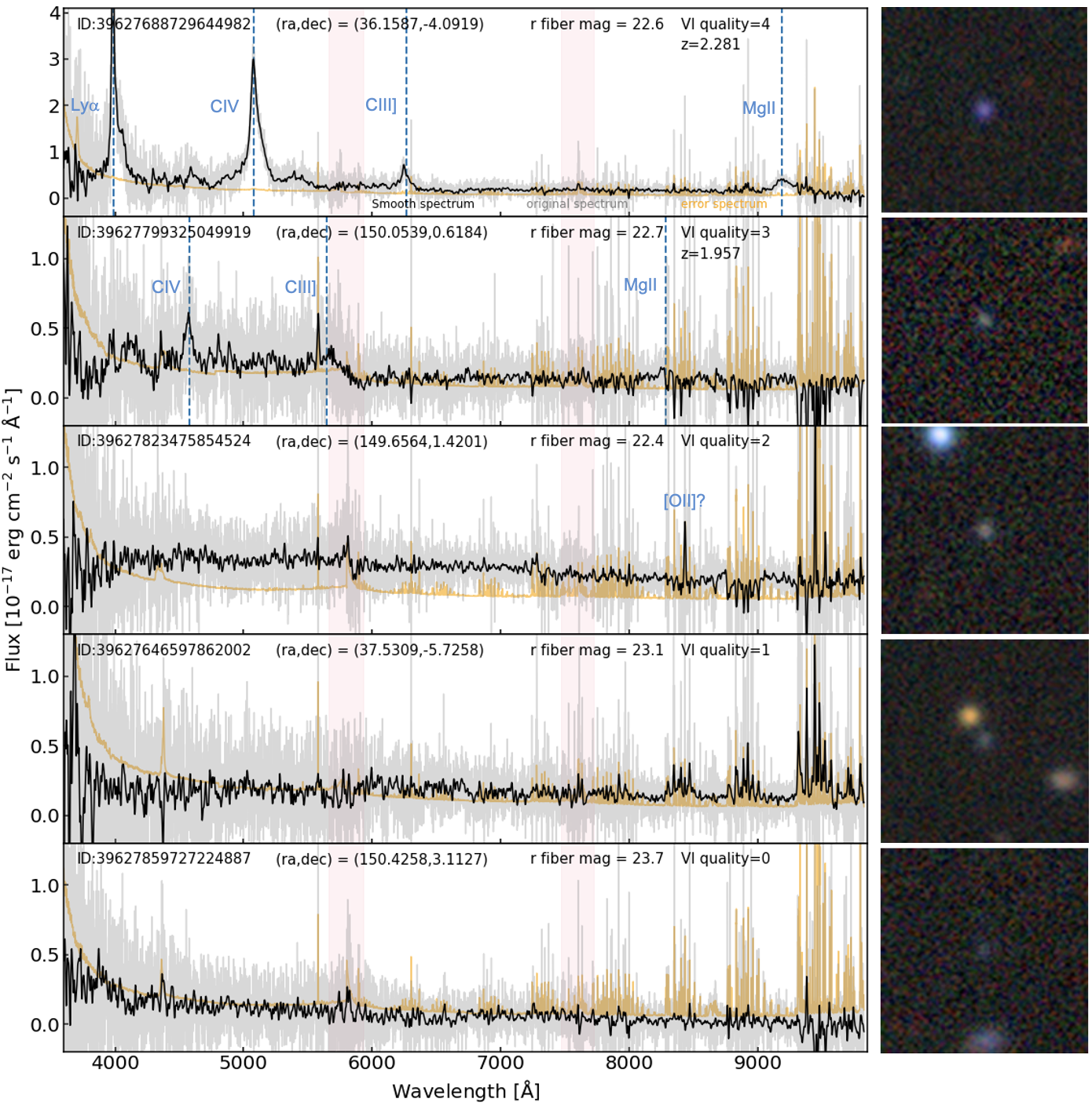}
\caption{Example DESI spectra illustrating the different VI quality classes (left) and thumbnail images ($18^{\prime\prime}\times18^{\prime\prime}$) centered on each target (right). Salient information for each target is plotted at the top of each spectrum including the VI quality flag. For each target both an unsmoothed (grey) and smoothed (black) spectrum are plotted along with the associated error spectrum (orange) and the overlapping wavelength ranges for the three different spectral arms (pale pink shaded regions). The most prominent emission lines identified in the two high-quality spectra (with VI quality flags of 3 and 4) are highlighted using blue vertical dashed lines as well as the potential identification of a single emission line ([{\sc o~ii}] at $z$~=~1.261) for the VI quality flag 2 target. We note that these visually inspected spectra were obtained with 10x longer exposures than those used in the main DESI survey.}
\label{fig:example_spectra}
\end{figure*}

\subsection{Visual-inspection approach}

Visual inspection of the optical spectra in DESI is crucial for several key reasons: the development and evaluation of the DESI spectroscopic pipeline processes, a quantitative assessment of the target-selection approaches, and the identification of physically interesting subsets from the selected-target populations. Given the scale of the main DESI survey ($>$~40 million spectroscopically identified targets) and the time-consuming nature of visual inspection, it is prohibitively expensive to visually inspect each spectrum.\footnote{As a rough guide it takes of order 1--2~mins to reliably VI an optical spectrum in the DESI quasar survey or $\approx$~5--10~years of continual person effort to VI all 2.8~million quasar spectra expected in the main DESI survey.} Visual inspection by its nature is also prone to human error and therefore requires each spectrum to be inspected by more than one person. Consequently, the VI efforts in DESI are primarily focused on assisting in the development of the pipeline and target-selection approaches to optimize both the effectiveness and quality of the spectroscopic outputs. 

The key focus of the VI is to assess the quality of each optical spectrum from the point of view of the measurement of a spectroscopic redshift to construct a ``truth table". Overall the same VI approach was adopted across both the QSO and galaxy survey components in DESI. However, due to the large variety of optical spectral types identified within the QSO survey (see \S3.1), a significant fraction of the VI effort within the QSO survey was also devoted to the spectroscopic classification of the selected targets. Consequently, this paper focuses more on the optical spectroscopic classifications of the targets than the galaxy VI paper (see Lan et~al.\ 2022) and, ultimately, calculates the performance of the QSO survey using only the spectrally classified QSOs rather than all  targets.

A summary of the different phases of VI for the QSO survey is provided in Table~\ref{table:VI_phases}. The initial two rounds were heavily focused on the development of the DESI spectroscopic pipeline and the refinement of the overall VI approach while the latter two rounds were more focused on quantifying the performance of the spectroscopic pipeline. Given the significant developments that occurred between the first two rounds and the later two rounds, the results presented in this paper are focused on the latter; i.e.,\ the deep-field VI and the sparse VI. The optical spectroscopic observations were also shorter in the first two VI rounds, with exposure times comparable to the main DESI survey ($\approx$~1,000~s) as compared to the $\approx$~3--10 times longer exposures for the deep-field and sparse VI. However, the basic VI approach whereby each target spectrum is assessed by more than one inspector and then merged by the QSO-survey VI lead (TMD for the first two rounds and DMA for the last two rounds), is common to both phases.\footnote{In the first two rounds every spectrum was evaluated by 3--4 visual inspectors while in the latter two rounds 2 visual inspectors evaluated each spectrum. In addition, the QSO-survey VI lead evaluated each spectrum where there was significant disagreement between visual inspectors.} 

Each target spectrum is evaluated using the \emph{Prospect} visualization tool.\footnote{\url{https://github.com/desihub/prospect}} A more extended description of \emph{Prospect} is provided in Lan et~al.\ (2022); however, the salient details relevant for an understanding of the QSO-survey VI are given here. \emph{Prospect} displays the unsmoothed optical spectrum and associated noise/error spectrum for each target spectrum; see Fig.~\ref{fig:example_spectra} for example spectra. The nine best-fitting redshift and spectral-template solutions calculated by the \emph{Redrock} spectral fitting code are also listed, along with the $\Delta\chi^2$ differences between each subsequent best-fitting solution. The visual inspector can smooth the spectrum and evaluate the spectral quality taking into account the overlapping regions of the three different spectral arms where breaks and discontinuities in the optical spectrum can occur. The visual inspector is then required to record:

\begin{itemize}
    \item Any problems and issues in the optical spectrum (e.g.,\ poor sky subtraction; significant breaks and discontinuities within a spectral arm).
    \item An assessment of the optical spectral classification (QSO, galaxy, or star) on the basis of the shape of the continuum emission and the identified emission and absorption lines. Targets are primarily classified as a QSO from the detection of broad permitted emission lines, strong non-stellar continuum, and/or prominent narrow high-excitation emission lines. Galaxies are identified based on the detection of low-excitation narrow emission and absorption features while stars are primarily identified from the detection of a stellar continuum with associated absorption lines at $z\approx$~0.
    \item The identification of physically interesting features or spectral sub types (e.g.,\ extreme absorption or emission features; identification of AGN spectral features within a galaxy spectrum; unusual optical continuum shapes; identification of a Broad Absorption Line Quasar (BALQSO), damped Ly~$\alpha$ system, blazar, or two objects contributing to the spectrum).
    \item An assessment of the spectroscopic redshift. The spectroscopic redshift is manually calculated by the visual inspector within {\it Prospect} by aligning the spectral features to the observed-frame wavelengths of potential emission and absorption lines, overlaid on the optical spectrum. The spectroscopic redshift can be refined in this way to 4 decimal places using a zoomed-in view of the spectrum if necessary.
    \item An assessment of the quality of the optical spectroscopic redshift using a numerical code from 0--4, defined as the VI quality class and described in more detail below.
\end{itemize}

The VI quality class indicates the reliability of the optical spectroscopic redshift measurement, as assessed by the visual inspector. The reliability of a spectroscopic redshift is related to more than just the signal-to-noise ratio of the spectrum and also depends on the spectral shape, strength of spectral features (i.e.,\ emission and absorption lines), and the redshift of the target (since the redshift will dictate which spectral features are present in the observed wavelength range).
Example spectra to illustrate the five different VI quality classes are provided in Fig.~\ref{fig:example_spectra} and associated qualitative descriptions are given below.

\vspace{0.2cm}
\noindent {\bf VI quality class 4:} the secure identification of two or more spectral features (i.e.,\ emission or absorption lines) indicating a confident spectroscopic redshift. Thanks to the high spectral resolution of the DESI data, the [{\sc o ii}]$\lambda$3727,3729 doublet can often be spectrally resolved, yielding a confident spectroscopic redshift even in the absence of no other identified features (see Fig.~3 of Lan et~al.\ 2022 for an example).

\vspace{0.2cm}
\noindent {\bf VI quality class 3:} the identification of one secure spectral feature plus additional weak spectral features indicating a probable spectroscopic redshift.

\vspace{0.2cm}
\noindent {\bf VI quality class 2:} the identification of one spectral feature indicating a possible spectroscopic redshift.

\vspace{0.2cm}
\noindent {\bf VI quality class 1:} continuum identified but no clear identified spectral features indicating an unreliable redshift.

\vspace{0.2cm}
\noindent {\bf VI quality class 0:} a weak or absent signal indicating a problem or issue with the optical spectrum.
\vspace{0.2cm}

The VI lead merges the results from all of the visual inspectors to determine the spectroscopic redshift, VI quality class, and spectral type for each target. We calculate the spectroscopic redshift difference between visual inspectors using the following basic formalism\
\begin{equation}
    dz_{\rm A,B}=\frac{|z_{\rm A}-z_{\rm B}|}{(1+z_{\rm A})},
    \label{eq:dz_vi}
\end{equation}
where in this case $dz_{\rm A,B}$ refers to the difference between the two VI redshifts, $z_{\rm A}$ and $z_{\rm B}$, and which we define as $dz_{\rm VI}$.

A mean spectroscopic redshift is calculated when there is good agreement between the visual inspectors ($dz_{\rm VI}<0.0033$) and a mean VI quality class is calculated when the quality class differs by $\le 1$ between visual inspectors.

For targets where there is a significant difference between the visual inspectors in terms of the spectroscopic redshift ($dz_{\rm VI}\ge0.0033$), the VI quality class (a difference of more than 1), or the selected spectral type (more than one spectral type selected), the VI lead uses \emph{Prospect} to re-evaluate the optical spectrum and manually determine the target redshift, quality class, and optical spectral type. Overall, for the QSO survey, the VI lead re-evaluates $\approx$~20\% of the optical spectra. 

A spectrum is considered high quality with a robust spectroscopic redshift when the VI quality class is $\ge$~2.5 (referred to hereafter as VI~$\ge$~2.5), calculated either from the average quality class of the visual inspectors or by the VI lead, when the spectrum is re-evaluated. Otherwise, the spectrum is considered low quality with an unreliable spectroscopic redshift; i.e.,\ VI~$<2.5$.

%\tablenum{A1}
\begin{table}
\caption{Tiles used in the deep-field visual inspection}
\label{table:VI_tiles_deep} 
\hskip-0.6cm\begin{tabular}{cccc}
\hline
TILEID & Position & $t_{\rm exp}$ & Common name\\
& (RA, DEC) [deg] & [s] & \\
\hline\hline
80605 & (36.448, -4.601)  & 7020 & XMM-LSS \\
80607 & (106.740, 56.100) & 9210 & Lynx \\
80609 & (150.120, 2.206) & 8300 & COSMOS \\
%\hline
%All & & & & 3761 \\
\hline
%\tablecomments{Note the exposure time corresponds to the median tile dark-time exposure.}
\end{tabular}
\tablecomments{The columns show the identification number of each DESI tile, the right ascension and declination of the tile center, the median dark-time exposure time, and the common names typically adopted for these well-studied fields.}
\end{table}

%\tablenum{A2}
\begin{table}
\caption{Tiles used in the sparse visual inspection}
\label{table:VI_tiles_sparse} 
\begin{tabular}{ccc}
\hline
TILEID  & Position & $t_{\rm exp}$\\
&  (RA, DEC) [deg] & [s]\\
\hline\hline
80620 &  (144.000, 65.000) & 6920 \\
80622 &  (155.000, 32.325) & 5390 \\
80669 &  (38.000, 0.500) & 4550 \\
80673 &  (85.500, -20.200) & 3540 \\
80674 &  (87.000, -23.200) & 2750 \\
80675 &  (98.500, 44.500) & 4120 \\
80676 &  (97.500, 47.700) & 4810 \\
80677 &  (104.500,	36.000) & 3550 \\
80678 &  (102.500,	39.000) & 4880 \\
80679 &  (111.000, 41.500) & 3470 \\
80680 &  (111.000, 44.800) & 6050 \\
80681 &  (115.000, 32.375) & 4800 \\
80682 &  (115.000, 32.375) & 4420 \\
80683 &  (116.000,	15.500) & 3890 \\
80684 &  (114.500, 18.400) & 3890 \\ 
80685 &  (120.000, 34.000) & 5380 \\
80686 &  (124.000, 34.300) & 3540 \\
80688 &  (130.700,	22.300) & 5560 \\
80690 &  (135.000,	32.375) & 6370 \\
80692 &  (139.000,	32.375) & 5620 \\
80693 &  (135.000, 83.000) & 5250 \\
80694 &  (162.000, 83.000) & 3650 \\
80699 &  (155.000,	32.375) & 4130 \\
80700 &  (159.000,	32.375) & 4800 \\
80707 &  (192.900, 27.100) & 5860 \\
80711 &  (213.000, 51.450) & 5280 \\
80712 &  (217.000, 53.550) & 3060 \\
\hline
\end{tabular}
\tablecomments{The columns show the identification number of each DESI tile, the right ascension and declination of the tile center, and the median dark-time exposure time.}
\end{table}

\subsection{Visual inspection data}

The data analysed in this paper is taken from the last two VI phases when the pipeline process was mature and stable: the deep-field and sparse VI (see Table~\ref{table:VI_phases}). Overall, $\approx$~6000 spectra were visually inspected in these latter VI phases. The exposures for the optical spectra from these datasets are $\approx$~3--10 times deeper than the nominal main DESI depth; see Tables~\ref{table:VI_tiles_deep} \& \ref{table:VI_tiles_sparse} for the depth of each tile. The higher S/N provided by these deep spectra allow for more reliable VI results and, consequently, more accurate truth tables. Since these deep spectra were obtained not from a single very long exposure but from the combination of many shorter exposures, they also allow for the construction of multiple shallower target spectra to replicate the quality of the spectra at the $\approx$~1000~s depth of the main 5~year survey, an aspect that we exploit to calculate the overall performance of the main DESI survey; see \S4.

\subsubsection{Deep-field VI datasets}

The focus of the deep-field VI was to undertake a complete visual inspection of the three deepest SV tiles observed and processed within the Blanc data assembly.\footnote{The data assembly refers to the DESI internal data release. With the exception of the first data assembly (SV0), all subsequent data assemblies have been named, in alphabetical order, after mountain ranges; see Table~\ref{table:VI_phases}. Consequently, Blanc and Cascades refer to the 3rd and 4th data assemblies, respectively. The most recent data assembly relevant for this paper is Fuji.} The key details of these three tiles are provided in Table~\ref{table:VI_tiles_deep} along with their more common field names: the XMM-LSS, Lynx, and COSMOS survey fields. Overall, all targets in these fields which met the SV1 quasar-target selection were visually inspected. The majority (3671) of the deep-field targets were visually inspected using  spectra from the Blanc data assembly. However, we also visually inspected the following spectra using the most-recent Fuji data assembly: (1) 18 additional targets which fell on fibers initially classified as ``bad" in the Blanc data assembly and (2) all 259 border-line low-quality targets with VI~=~2.0--2.9, identifying the majority as high-quality spectra (VI~$>$~2.5) with the Fuji data assembly.

The deep-field VI dataset provides the most complete evaluation of the overall quality of the optical spectra in the DESI quasar survey. It allows for a quantitative assessment of the standard \emph{Redrock} pipeline process in the measurement of optical spectroscopic redshifts and the identification of the optical spectral classes within the quasar survey.

\begin{figure}
\center
\includegraphics[width=0.4\textwidth]{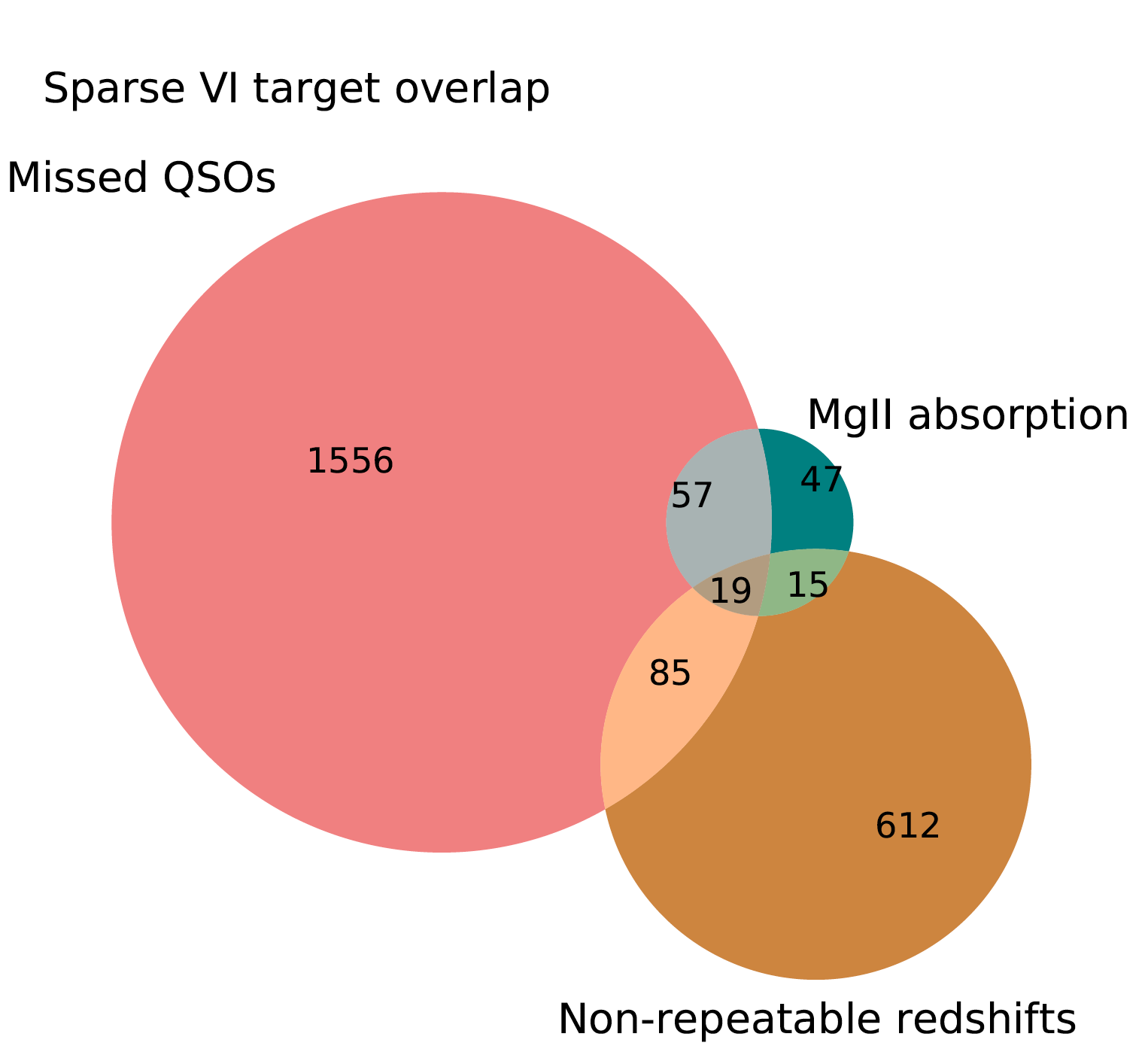}
\includegraphics[width=0.4\textwidth]{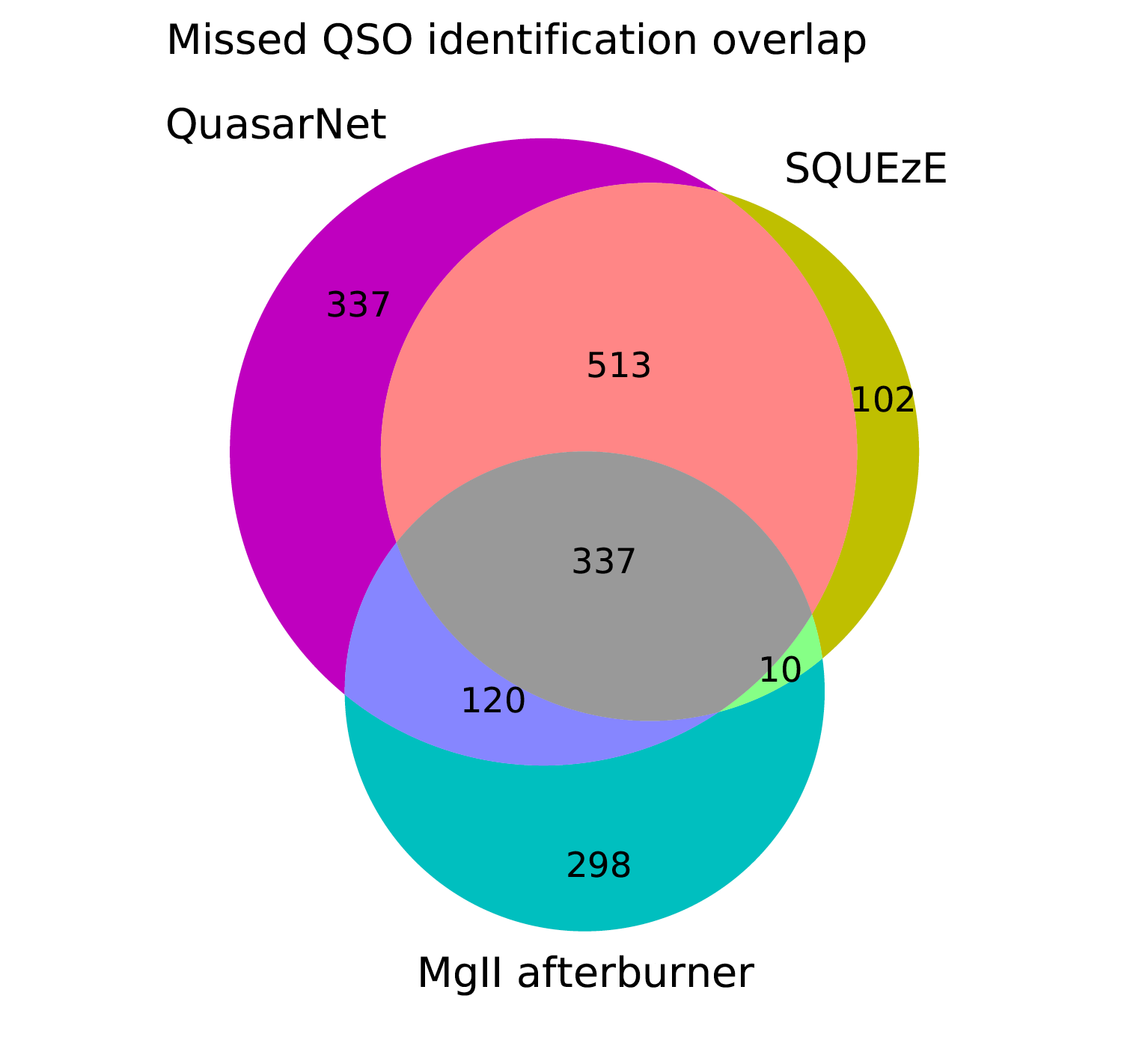}
\caption{Venn diagrams showing (top) the overlap between the missed QSO sample, the Mg~II absorption systems, and the targets with non-repeatable redshifts and (bottom) the overlap in the selection of missed QSOs between the three afterburner selections across the sparse VI fields.}
\label{fig:venn}
\end{figure}

\subsubsection{Sparse VI datasets}

The three fields inspected in the deep-field VI (described in \S2.3.1) only comprise $\approx$~10\% of the SV tiles in the quasar survey. To complement the deep-field VI we also visually inspected targets in the other 27 SV tiles observed and processed within the Cascades data assembly. However, unlike the deep-field VI, we focused our efforts on visually inspecting key subsets of the selected quasar targets to either improve the source statistics of comparatively rare target subsets identified in the deep-field VI or to test potential issues in the identification pipeline. Consequently, the visual inspection of these fields is referred to as the ``sparse VI". The key details of these 27 tiles are provided in Table~\ref{table:VI_tiles_sparse} and a breakdown of the target selections for the sparse VI is given in Table~\ref{table:sparse_subsets} and described below.

For the quasar survey the main focus of the sparse VI was to calculate the performance of several independent quasar-identification approaches (QuasarNet, SQUEzE, and Mg~II afterburner); these approaches are referred to generically here as ``afterburners" since they are performed on the DESI spectroscopy following the standard \emph{Redrock} spectroscopic pipeline. As demonstrated from the deep-field VI in \S3, while the spectral classification of a quasar by the standard \emph{Redrock} pipeline is highly reliable, it misses a non-negligible fraction of genuine quasars, the majority of which we can recover through the afterburner approaches; see \S3.2 and the brief afterburner descriptions below. We generically refer to QSOs missed by \emph{Redrock} as ``missed QSOs".

QuasarNet (QN; Busca \& Balland 2018) utilizes a deep convolutional neural network (CNN) classifier with multiple layers of convolution to identify emission lines and to calculate the most likely target redshift. SQUEzE (SQ; Perez-Rafols et~al.\ 2020a,b) also adopts a machine-learning approach but uses a random-forest classifier to assess the identification of emission peaks and to calculate the most likely target redshift. The performance of QN and SQ has been previously evaluated using the visually inspected spectra from the SDSS BOSS quasar survey (Farr, Font-Ribera, \& Pontzen 2021) to assess the utility of these approaches in the quasar identification in the DESI quasar survey. By comparison, the Mg~II afterburner searches for significant broad Mg~II emission at the best-fitting \emph{Redrock} redshift for systems spectrally classified as galaxies (see \S6.2 of Chaussidon et~al.\ 2022). It therefore differs from QN and SQ in not independently calculating the target redshift but switches the spectroscopic classification from galaxy to quasar when probable broad Mg~II emission is detected. 

\begin{table}
\caption{Target subsets explored in the sparse VI}
\label{table:sparse_subsets}
\hskip1.0cm\begin{tabular}{cc}
\hline
Target selection & N \\
\hline\hline
QuasarNet & 1307 \\
SQUEzE & 962 \\
MgII afterburner & 765 \\
\hline
Unique missed QSOs & {\bf 1717} \\
\hline
MgII absorption systems & 138 \\
Non-repeatable redshifts & 731 \\
\hline
Unique targets & {\bf 2391} \\
\hline
\end{tabular}
\tablecomments{Each row shows the number of targets visually inspected for each target-selection approach in the sparse VI. The total number of unique targets from the missed QSO selection approaches is highlighted in bold in addition to the total number of unique targets visually inspected across all target-selection approaches; the number of overlapping targets between each selection approach is shown in Fig.~\ref{fig:venn}.}
\end{table}

In selecting targets for the missed QSO component of the sparse VI we required an optical magnitude of $r<23.2$, the non-identification of a quasar from the best-fitting \emph{Redrock} template solution from the Cascades data assembly, and a minimum confidence threshold depending on the afterburner approach: $QN\_C\_LINE\_BEST>0.95$ for QN, $PROB>0.4$ for SQ, or a $\Delta\chi^2>16$ improvement in the fitted spectra when broad Mg~II emission is included for the Mg~II afterburner. We chose these thresholds based on the results from the afterburner codes performed on the deep-field VI to achieve a balance between selecting genuine missed QSOs and providing a dataset for which to further assess the reliability of the afterburner approaches. Although the three afterburners differ in their identification approaches there is a reassuringly high level of overlap between the selected targets; see the target-selection approach breakdown in Table~\ref{table:sparse_subsets} and the Venn diagram in the lower panel of Fig.~\ref{fig:venn}. 

In addition to the missed QSOs we also visually inspected (1) a sample of $r<23.2$~mag candidate Mg~II absorption systems and (2) targets where the best-fitting \emph{Redrock} spectral template--redshift solutions differed significantly between the coadded spectroscopy and shallower nominal main DESI survey depth data. The latter are referred to as ``non-repeatable redshifts" and selected when
\begin{equation}
            \Delta\chi^2 ({\rm short})>20\ \&\ \Delta\chi^2 ({\rm long})>100\ \&\ dz>0.05,
            \label{eq:dz_vi}
        \end{equation}
\noindent where $dz$ refers to the relative redshift difference between the measured redshifts of individual short ($>700$~s) and coadded long ($>3000$~s) exposure spectra, and $\Delta\chi^2$ refers to the difference in reduced $\Delta\chi^2$ between the best-fitting and 2nd best-fitting \emph{Redrock} spectral template solutions. See Eqn.~1 for the basic formalism for calculating $dz$ using the short and long exposure redshifts.

The VI of the Mg~II absorption systems was undertaken to assist in the development of a Mg~II absorption-line selection tool and will be presented in Napolitano et~al.\ (in prep). The VI for the non-repeatable redshifts were taken to analyse the spectroscopic pipeline and are not further investigated in this paper. For completeness, we summarise the number of selected targets from these target-selection approaches in Table~\ref{table:sparse_subsets} and illustrate their modest overlap with respect to the missed QSOs in the top panel of Fig.~\ref{fig:venn}.

\section{Visual inspection: basic results}

In this section we present the basic results from the visual inspection of the quasar survey using the deep co-added SV spectra, focusing on the deep-field VI (\S3.1) and the missed QSO targets in the sparse VI (\S3.2); we defer our assessment of the expected performance from the shallower main-survey spectra to \S4. To motivate the missed QSO results, in \S3.1 we also quantify the effectiveness of the standard \emph{Redrock} spectroscopic pipeline in the identification of QSOs. In \S3.3 we use the VI results to test the modified pipeline approach presented in Chaussidon et~al.\ (2022). When comparing our VI results to those from the \emph{Redrock} spectroscopic pipeline we have used the redshifts and spectral classifications from the most-recent Fuji data assembly to ensure the inclusion of the latest updates to the spectroscopic pipeline and data-reduction procedures.

\begin{figure*}
\center
\includegraphics[width=0.9\textwidth]{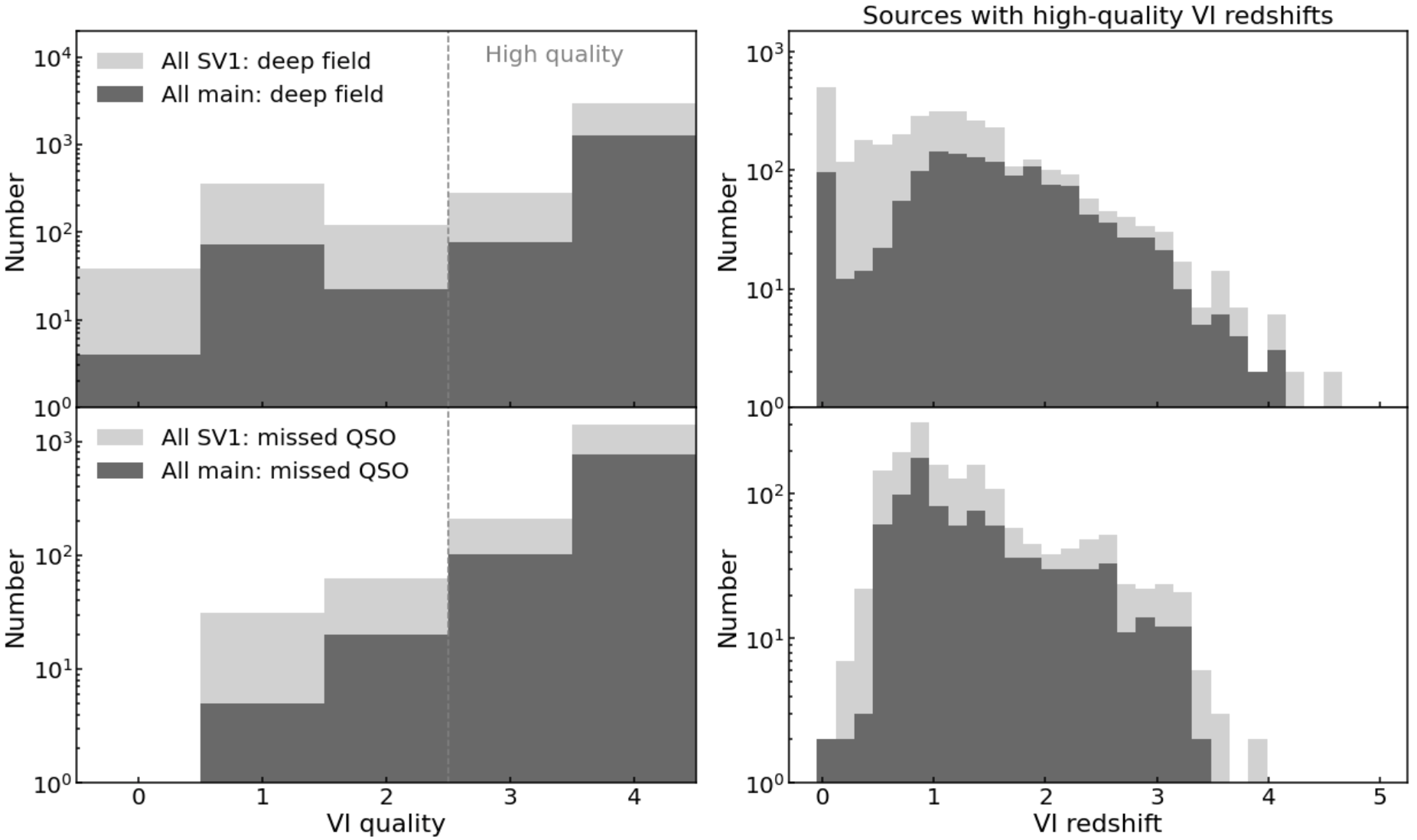}
\caption{Distribution of (left) VI quality flags and (right) high-quality VI redshifts for the deep-field VI (top) and missed QSOs in the sparse VI (bottom). Targets meeting the SV1 and main selection are indicated in pale and dark grey, respectively. The dashed line indicates the threshold required for a high-quality redshift (VI~$\ge2.5$).}
\label{fig:VI_quality}
\end{figure*}

\begin{figure}
\center
\includegraphics[width=0.4\textwidth]{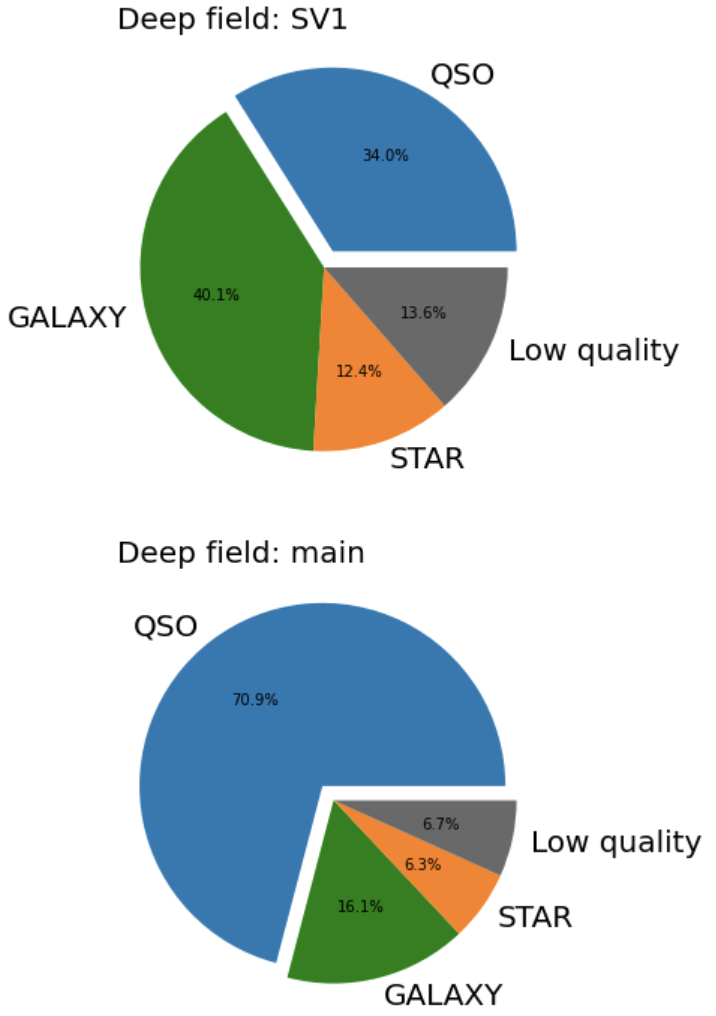}
\caption{Pie chart showing the distribution of high-quality optical spectral classifications and low-quality spectra from the deep-field VI for targets meeting the SV1 (top) and main (bottom) target selections.}
\label{fig:pie}
\end{figure}

\begin{table*}
\caption{Results from the deep-field VI and missed-QSO sparse VI}
\label{table:results_missed_sparse}
\centerline{
\hskip-2.5cm\begin{tabular}{c|c|cc|cccc}
\hline
VI dataset & Selection & N & High quality & High quality & High quality & High quality & Low quality\\
 & & (all) & (all)    & (QSO)        & (GALAXY)        & (STAR)       & \\
\hline\hline
Deep field & SV1 & 3779 & 3266 (86.4\%)  & 1283 (34.0\%) & 1516 (40.1\%) & 467 (12.4\%) & 513 (13.6\%)\\
%\hline

 & main & 1455 & 1357 (93.3\%)  & 1032 (70.9\%) & 234 (16.1\%)  & 91 (6.3\%)  & 98 (6.7\%)\\
\hline
Missed QSO & SV1 & 1717 & 1624 (94.6\%) & 1489 (86.7\%) & 133 (7.7\%) & 2 (1.2\%) & 93 (5.4\%)\\
 & main & 899 & 874 (97.2\%)   & 850 (94.5\%)  & 22 (2.4\%) & 2 (2.2\%) & 25 (2.8\%)\\
\hline
\end{tabular}}
\tablecomments{The VI results are presented as a function of the identified optical spectral class (QSO; GALAXY; STAR), split into high quality (VI~$\ge$~2.5) and low quality (VI~$<2.5$) spectra. The results are calculated for targets from the deep field or missed QSO VI datasets meeting either the SV1 (top row) or main (bottom row) target selections. The corresponding percentages are calculated from all visually inspected targets meeting a given target-selection approach.}
\end{table*}

\begin{figure*}
\center
\includegraphics[width=0.49\textwidth]{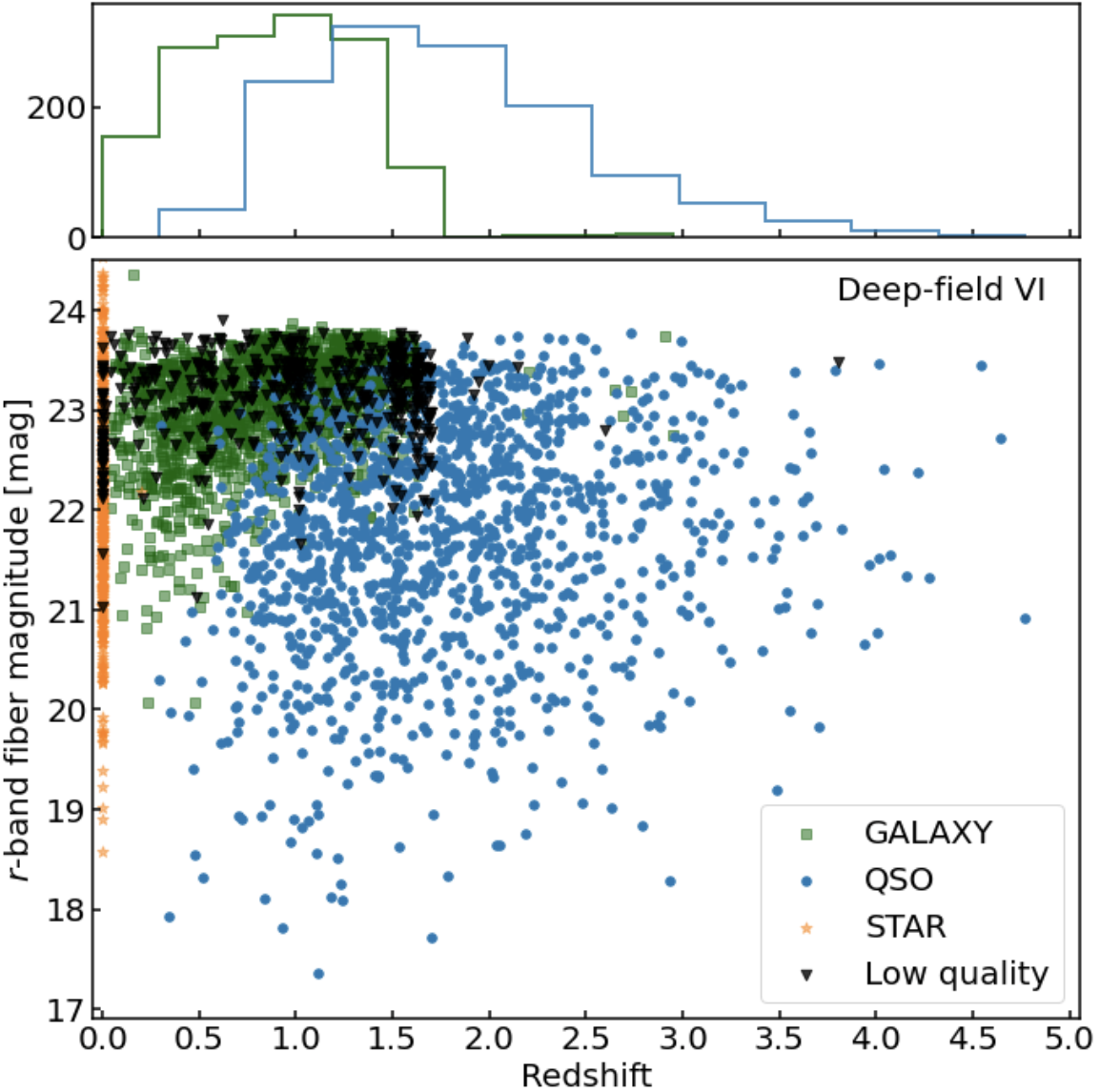}
\includegraphics[width=0.49\textwidth]{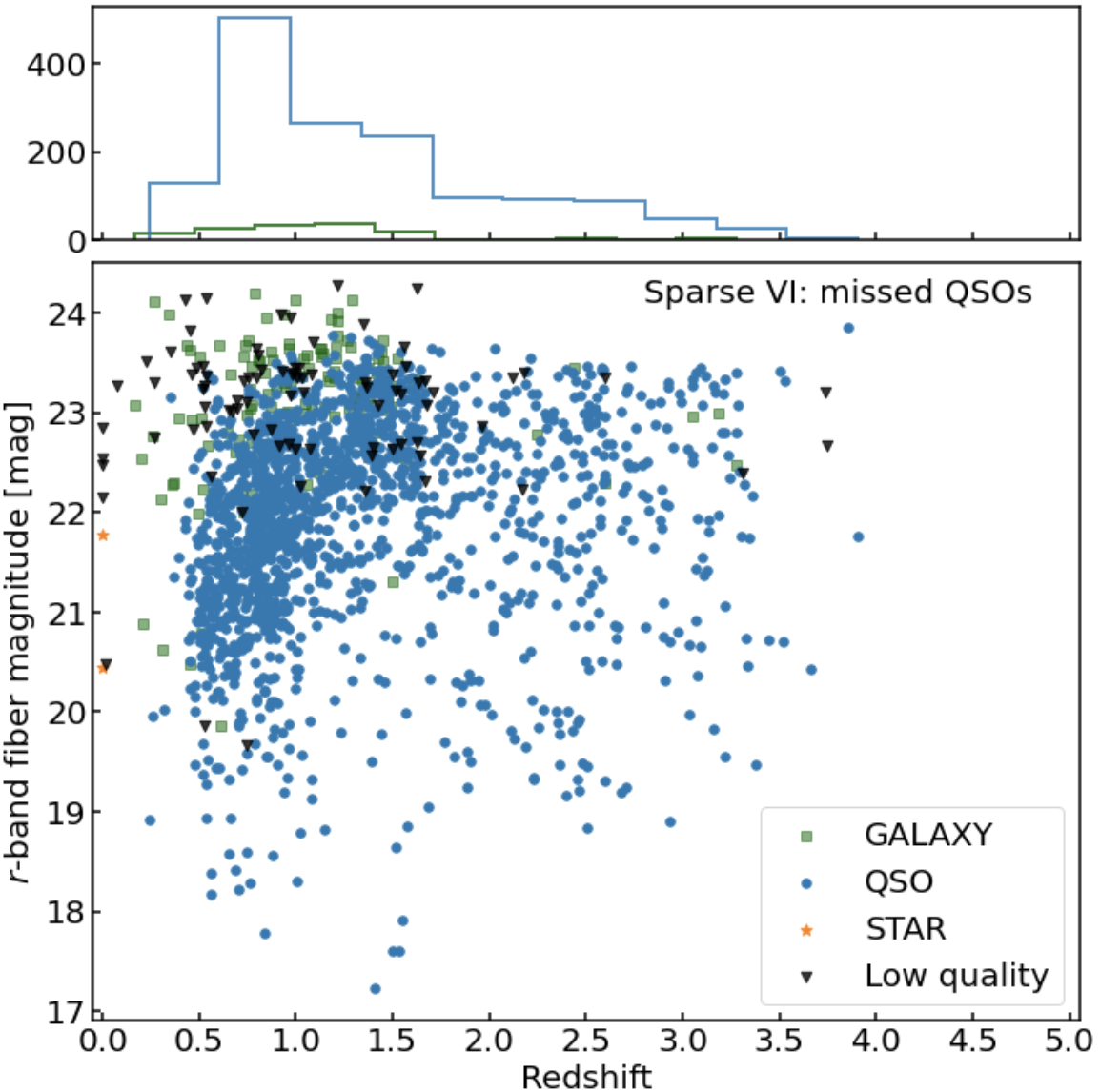}
\caption{$r$-band fiber magnitude versus redshift for (left) the deep-field VI (see section 2.3.1) and (right) the missed QSOs from the sparse VI (see section 2.3.2); the redshift is taken from the VI for the high-quality spectra and from \emph{Redrock} for the low-quality spectra. The target classifications are based on the visually inspected optical spectral type for the high-quality spectra (VI~$\ge$~2.5): QSO (blue circle), galaxy (green square), star (orange star). The low-quality (VI~$<2.5$) identifications are plotted as black triangles. The top panels show the redshift distributions for the high-quality QSOs (blue) and the galaxies (green).}
\label{fig:r-band}
\end{figure*}

\begin{table*}
\caption{Pipeline-identification results in the deep fields}
\label{table:redrock_results_deep}
\hskip-2.0cm\begin{tabular}{c|c|ccc|ccc}
\hline
 &  &  \multicolumn{3}{c}{Standard pipline} & \multicolumn{3}{c}{Modified pipline} \\
\hline
Selection & Spectral & N & High quality  & High quality & N & High quality  & High quality\\
&  class  &   & (correct identification) & (missed) &  & (correct identification) & (missed) \\
\hline\hline
%SV1 & All  & 3761 & 3204 (85.2\%) & \dots \\
SV1 & QSO  & 1135 & 1102 (97.1\%) & 181 (14.1\%) & 1345 & 1262 (93.3\%) & 21 (1.6\%) \\
    & GALAXY & 2175 & 1503 (69.1\%) & 13 (0.9\%) & 1971 & 1467 (74.4\%) & 49 (3.2\%) \\
    & STAR & 469  & 414 (88.3\%) & 53 (11.3\%) & 463 & 414 (89.4\%) & 53 (11.3\%) \\
\hline
%Main & All  & 1447 & 1337 (92.7\%) & \dots \\
Main & QSO  & 928 & 923 (99.5\%) & 109 (10.6\%) & 1042 & 1025 (98.4\%) & 7 (0.7\%) \\
     & GALAXY & 447 & 232 (51.9\%) & 2 (0.9\%)  & 338 & 224 (66.3\%) & 10 (4.3\%) \\
     & STAR & 80 & 70 (87.5\%) & 21 (23.1\%)  & 75 & 70 (93.3\%) & 21 (23.1\%) \\
\hline
\end{tabular}
\tablecomments{The pipeline-identification results in the deep fields as a function of the best-fitting optical spectral class for targets meeting either the SV1 (top) or main (bottom) target selection. The results are split between those obtained using the standard \emph{Redrock} pipeline and the modified pipeline. The number of high-quality spectra (VI~$\ge$~2.5) with the same optical spectral class as that found from the VI is provided; the corresponding percentage is calculated from all targets with the same spectral class. In addition the number of high-quality targets identified from the VI but missed by the pipelines are listed; the corresponding percentage is calculated from the total number of visually inspected targets with the same optical spectral class as given in Table~\ref{table:results_missed_sparse}.}
\end{table*}

\subsection{Deep-field VI}

In Fig.~\ref{fig:VI_quality} (top panels) we show the distribution of VI quality classes and high-quality (VI~$\ge$~2.5) VI redshifts from the deep-field VI for both the SV1 and main target selections. The basic results from the deep-field VI are also presented in Table~\ref{table:results_missed_sparse}. The majority of the optical spectroscopy is of high quality (VI~$\ge$~2.5), yielding reliable spectroscopic redshifts and optical spectral classifications; we also note that there are $\approx$~10x more VI~=~4 spectra than VI~=~3 spectra.

A larger fraction of targets ($\approx$~93\%) that meet the conservative main-target selection have high-quality spectra than found for the fainter and more liberal SV1 selection ($\approx$~86\%). The high-quality redshift distributions peak around $z\approx$~1 for both the main and SV1 target selections and includes all spectroscopic classes (QSOs; galaxies; stars); indeed, the significant peak at $z\approx0$ is due to a non-negligible fraction of stars. At $z>1.6$, the redshift threshold uniquely traced by the QSO survey within DESI, there are modest differences between the SV1 and main target selection approaches, demonstrating that the more conservative main target selection provides comparable high-redshift source statistics to the liberal SV1 selection. The excess number of $z<1.6$ targets in the SV1 selection are a combination of galaxies, lower-redshift QSOs, and stars.

In Fig.~\ref{fig:pie} the distributions of optical spectral class for both the main and SV1 target selections are shown, which are further quantified in Table~\ref{table:results_missed_sparse}. The main selection is significantly more efficient than SV1 at selecting QSOs ($\approx$~34\% for SV1 and $\approx$~71\% for main). However, a significant fraction of targets are classified as either galaxies or stars for both selections; indeed, for SV1 galaxies are more dominant than QSOs. Such a large diversity in optical spectral classification is not seen in the DESI galaxy surveys (Lan et~al.\ 2022) and is largely due to the challenge in robustly selecting QSOs from stars and galaxies using only photometric data. At the faint optical magnitudes probed by DESI, distinguishing between galaxies and QSOs becomes more challenging than for shallower optical surveys such as the SDSS since the emission from the host galaxy can become as significant as the QSO for the lower luminosity (or more dust-reddened) systems identified. 

%Overall, the SV1 target selection yields $\approx$~24\% more high-quality QSOs than achieved from the main target selection but at the expense of selecting $\approx$~2.6 times more targets.

In Fig.~\ref{fig:r-band} (left) the $r$-band fiber magnitude versus redshift for the SV1 target selection in the deep-field VI is shown, with targets plotted on the basis of optical spectral class. The galaxies and QSOs have different redshift distributions with the galaxies predominantly at low redshift, with few at $z>1.5$, while the QSOs extend out to $z\approx$~5, peaking at $z\approx$~1.0--2.5. Most of the low-quality spectra have faint magnitudes ($r>23$~mags), at least partially explaining the significantly higher fraction of low-quality spectra in the fainter SV1 selection than for the main selection (i.e.,\ lower S/N data); see Table~\ref{table:results_missed_sparse}. However, the majority of the low-quality identifications occur at distinct redshift peaks and ranges. These redshift peaks are largely due to \emph{Redrock} erroneously identifying a noise feature (often due to poor sky subtraction) as an emission line; e.g.,\ the most prominent peak around $z\approx$~1.5--1.7 is due to \emph{Redrock} identifying [{\sc o~ii}] at $\approx$~9300--10000\AA\, where the data can be particularly noisy (see the VI~=~1 spectrum in Fig.~\ref{fig:example_spectra} for a good example).

The results presented so far are based solely on the VI. However, since the vast majority of the optical spectra in the DESI survey will not be visually inspected, it is instructive to compare the VI results to the output from the spectroscopic pipelines. In Table~\ref{table:redrock_results_deep} the pipeline results for both the SV1 and main target selections are shown, split as a function of the optical spectral class and sub divided between the standard and modified pipelines. Here we focus on the results from the standard \emph{Redrock} pipeline and defer comparisons with the modified pipeline to \S3.3. The standard \emph{Redrock} pipeline reliably identifies QSOs, with $\approx$~97\% (SV1) and $\approx$~99.5\% (main) visually confirmed as high-quality QSOs. A slightly lower $\approx$~88\% of \emph{Redrock}-identified stars are visually confirmed as high-quality stars, independent of target-selection approach, while by comparison $<69$\% of the \emph{Redrock}-identified galaxies are found to be high-quality identifications; indeed, the vast majority of the low-quality identifications reported in Table~\ref{table:results_missed_sparse} are classified by \emph{Redrock} as galaxies. However, despite the high reliability of \emph{Redrock}-identified QSOs, a non-negligible fraction of high-quality QSOs identified in the VI are missed by the standard \emph{Redrock} pipeline ($\approx$~14\% for SV1 and $\approx$~11\% for main). As shown in the following sub section, the majority of the high-quality QSOs missed by \emph{Redrock} can be recovered with the afterburners.

\begin{figure}
\center
\includegraphics[width=0.45\textwidth]{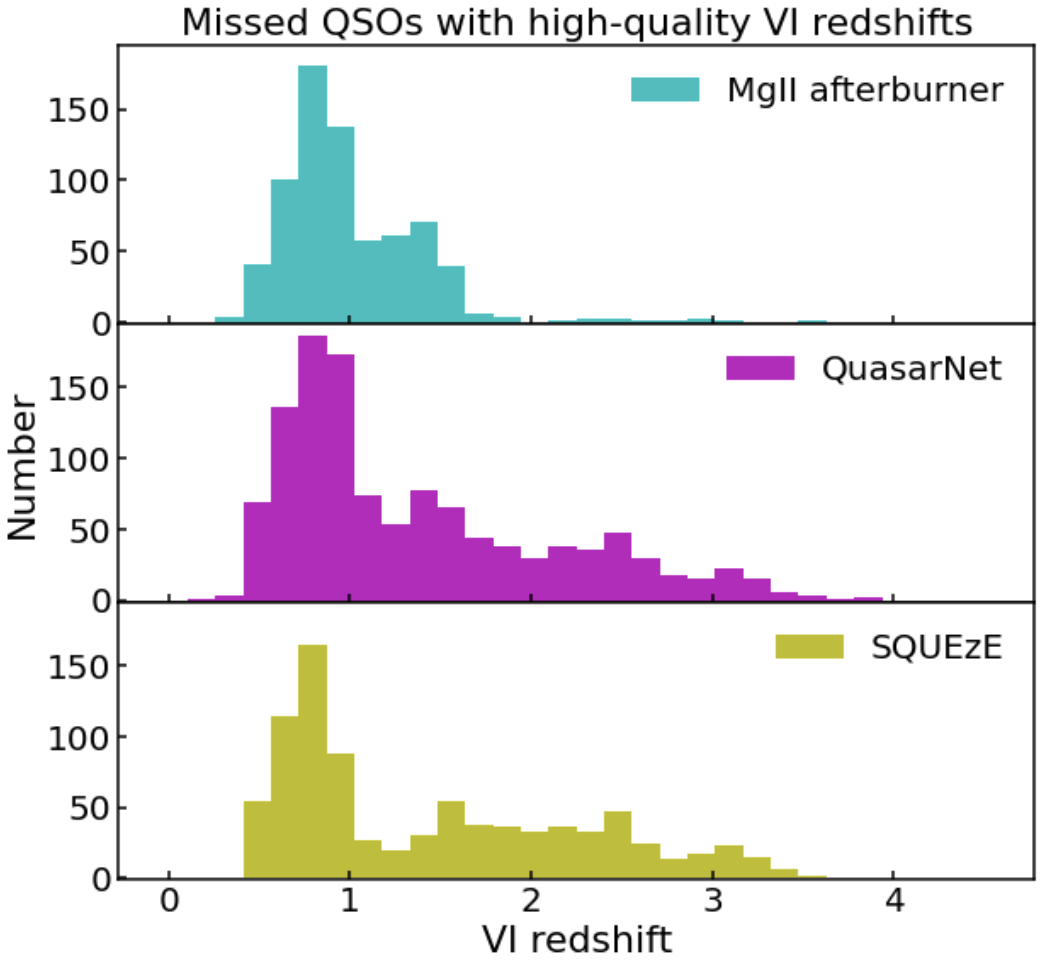}
\caption{Redshift distributions for high-quality QSOs from the missed QSOs in the sparse VI for (top) Mg~II afterburner selected systems, (middle) QN selected systems, and (bottom) SQ selected systems.}
\label{fig:zdist}
\end{figure}

\begin{table}
\caption{Missed-QSO sparse VI results for the afterburners}
\label{table:afterburner_reliability}
\centerline{
\hskip-1.5cm\begin{tabular}{c|c|cc}
\hline
 &  &  & High quality\\
Afterburner & Selection & N (all) & N (QSO) \\
\hline\hline
QuasarNet & SV1  & 1307 & 1182 (90.4\%)\\
%\hline
& Main & 744 & 718 (96.5\%)\\
\hline
SQUEzE & SV1  & 962 & 881 (91.6\%)\\
%\hline
& Main & 584 & 569 (97.4\%)\\
\hline
Mg~II & SV1  & 765 & 709 (92.7\%)\\
%\hline
& Main & 429 & 413 (96.3\%) \\
\hline
\end{tabular}
}
\tablecomments{The total number of targets for each afterburner approach and the number (and associated overall percentage) with high-quality QSO identifications (VI~$\ge$~2.5) are given for targets meeting either the SV1 (top) or main (bottom) target selection.}
\end{table}

\begin{figure}
\center
\includegraphics[width=0.47\textwidth]{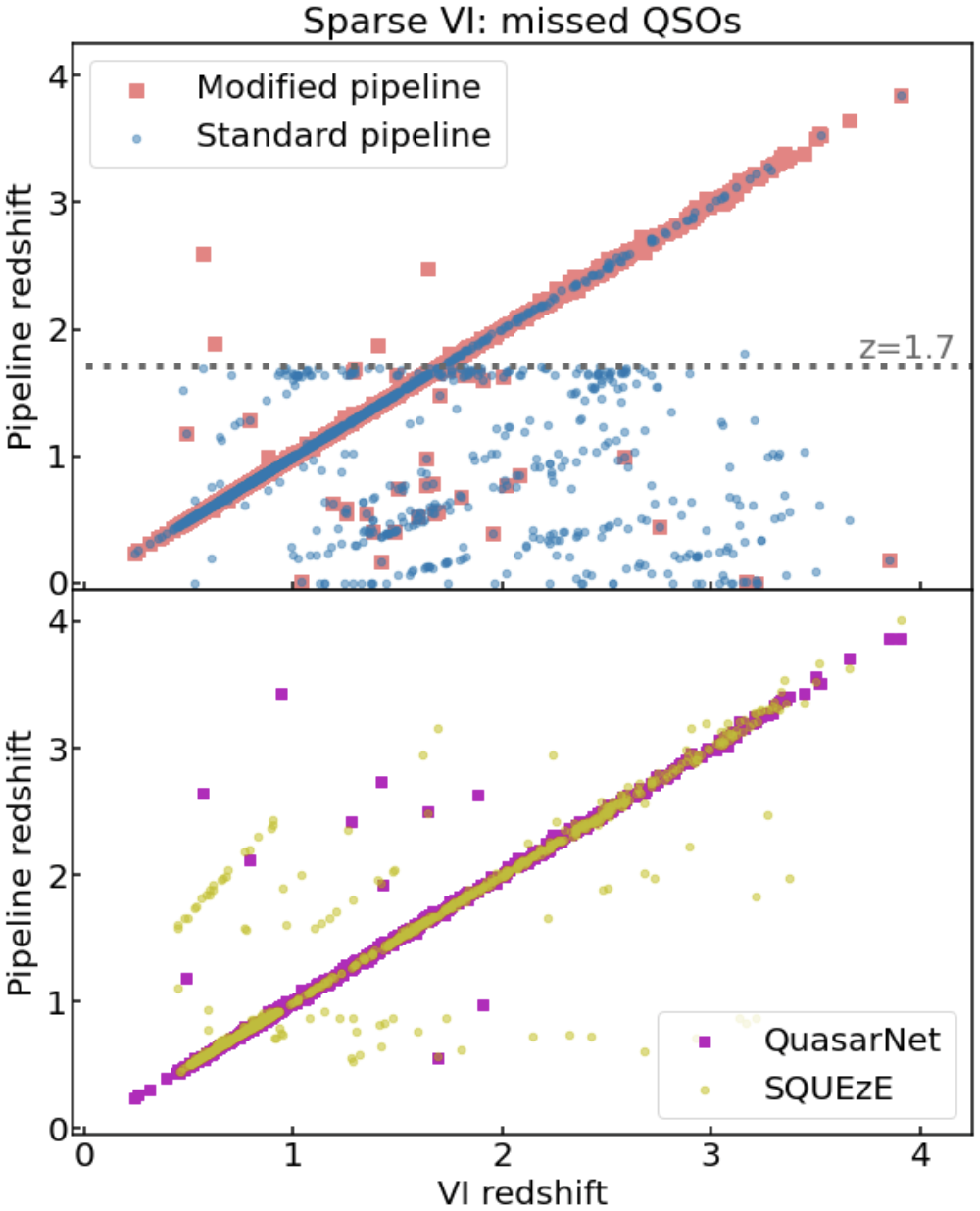}
\caption{Pipeline redshift versus VI redshift for the high-quality missed QSOs from the sparse VI. The plotted pipeline redshifts from (top) the standard \emph{Redrock} pipeline (blue) and modified pipeline (red) and (bottom) the QuasarNet (magenta) and SQUEzE (yellow) afterburners. We note that all missed QSOs are plotted for the standard and modified pipelines while, for QuasarNet and SQUEzE, only the missed QSOs identified by that afterburner are plotted.}
\label{fig:zdiff_missed}
\end{figure}

\subsection{Sparse VI: missed QSOs}

In Fig.~\ref{fig:VI_quality} (bottom panels) we show the distribution of VI quality classes and high-quality VI redshifts for the missed QSO selection from the sparse VI for both the SV1 and main target selections. The missed QSO VI results are also presented in Table~\ref{table:results_missed_sparse}. As for the deep-field VI, the majority of the optical spectra are high quality: $\approx$~95\% (SV1) and $\approx$~97\% (main) have VI~$\ge$~2.5. The even larger fraction of high-quality spectra, as compared to the deep-field VI, is likely due to the additional requirement that the targets exceed the confidence threshold required by at least one of the afterburners. Individually each of the afterburner approaches selects $>90$\% high-quality QSOs, which rises to $>96\%$ when focusing on the main-target selection, demonstrating the high reliability of the afterburners in identifying high-quality QSOs; see Table~\ref{table:afterburner_reliability}.

In Fig.~\ref{fig:r-band} (right) we show the $r$-band fiber magnitude versus redshift for the SV1 target selection for the missed QSOs, with targets plotted on the basis of optical spectral class. A significant difference is seen between the QSOs in the sparse VI and the deep-field VI (Fig.~\ref{fig:r-band}, left) with many of the former populating a clearly defined curved band across the $r$-band--redshift plane at $z\approx$~0.4--1.7, leading to a narrower overall redshift distribution (see Fig.~\ref{fig:VI_quality}). The primary origin of this $r$-band--redshift curved band is the identification of broad Mg~II emission at the redshift of a \emph{Redrock}-identified galaxy. This is illustrated in Fig.~\ref{fig:zdist}: the Mg~II afterburner targets are tightly distributed over $z=$~0.4--1.7. For these missed QSOs, \emph{Redrock} has measured the correct redshift but classified the target as a galaxy rather than a QSO.

However, the QuasarNet and SQUEzE afterburners have also identified a population of missed QSOs across a wider redshift range. For these systems, \emph{Redrock} failed to identify the QSO and also measured the wrong redshift. To illustrate these points we plot in Fig.~\ref{fig:zdiff_missed} (top) the redshift from the standard \emph{Redrock} pipeline versus the VI redshift: good agreement is found for many systems at $z<1.7$ but a significant fraction of targets clearly have catastrophic redshift failures.\footnote{The catastrophic redshift errors are primarily due to the limited spectral template range available for QSOs in the standard \emph{Redrock} pipeline, limiting the ability for \emph{Redrock} to reliably capture the spectral diversity across the QSO population. Indeed, as shown in \S5.1, the missed QSOs with redshift failures have redder optical spectra (due to larger host-galaxy contributions and dust reddening) than the standard \emph{Redrock} QSOs. The limited spectral template range for the QSOs is driven by the need for \emph{Redrock} to fit spectra across all of the galaxy surveys in addition to the QSO survey. Expanding the range of QSO templates in the standard \emph{Redrock} pipeline leads to degeneracies between the identification of red galaxies (such as those identified in the LRG survey) and red QSOs, which are the minority population, which would significantly compromise the overall performance of DESI.\label{footnote:catastrophicz}} However, as shown in Fig.~\ref{fig:zdiff_missed} (bottom), better redshift agreement is achieved when either the QuasarNet or SQUEzE afterburner redshift is adopted in preference to \emph{Redrock}, providing the foundations for the construction of the modified pipeline for the QSO survey (Chaussidon et~al.\ 2022), as described next. 

\subsection{Testing the effectiveness of the modified pipeline with the visual-inspection data}

On the basis of the VI results presented in \S3.1 the standard \emph{Redrock} pipeline reliably distinguishes QSOs from stars and galaxies with just a small fraction of \emph{Redrock}-identified QSOs with low-quality spectra ($\approx$~4\% for SV1 and $\approx$~1\% for main). A non-negligible fraction of high-quality QSOs ($\approx$~14\% for SV1 and $\approx$~11\% for main) are missed by the standard \emph{Redrock} pipeline but, as demonstrated in \S3.2, we can reliably recover the majority using the afterburners. Consequently, a combination of the QSO identification approaches from both \emph{Redrock} and the afterburners should lead to a larger selection of high-quality QSOs. This is the key motivation for the development of a modified pipeline to optimize the selection of QSOs (Chaussidon et~al.\ 2022), commonly referred to as ``QSO maker". In the following analyses we use the VI results to quantify the performance of the modified pipeline and compare these results to the standard \emph{Redrock} pipeline.

Below we provide a summary of the key steps in the modified pipeline and refer the reader to the flowchart in Fig.~9 of Chaussidon et~al.\ (2022) for an illustration. Essentially, targets are classified as QSOs when identified by either the standard \emph{Redrock} pipeline, the Mg~II afterburner, or QuasarNet, adopting the same afterburner confidence thresholds used for the sparse VI missed-QSO selection (see \S2.3.2). \emph{Redrock} is then used to recalculate the spectroscopic redshift when either (1) QuasarNet identifies a QSO unidentified by the Mg~II afterburner or (2) when QuasarNet identifies the target as a QSO and the \emph{Redrock} redshift disagrees with the QuasarNet redshift (i.e.,\ $dz>0.05$; see Eqn.~1 for the basic formalism of $dz$). To improve the final redshift measurement, the QuasarNet redshift is used as an input prior to \emph{Redrock} with a tight top-hat redshift distribution of $dz$~=~0.05, and only QSO template solutions are adopted in the redshift fitting. 

To quantify the effectiveness of the modified pipeline for the identification of QSOs, we compare to the standard \emph{Redrock} pipeline and the VI in the deep fields; see Table~\ref{table:redrock_results_deep}. The modified pipeline identifies a significantly larger number of QSOs for both the main and SV1 selections, the vast majority of which are high quality ($\approx$~93\% for SV1 and $\approx$~98\% for main). Furthermore, almost all of the QSOs missed by the standard \emph{Redrock} pipeline are reliably recovered by the modified pipeline (just 1--2\% of the high-quality QSOs are missed), providing a $>10$\% increase in the number of high-quality QSOs. The overall number of QSOs with low-quality spectra is increased by factor $\approx$~3 compared to the standard \emph{Redrock} pipeline but they only account for $\approx$~2\% of the modified-pipeline QSOs for the main target selection ($\approx$~7\% for SV1).

The modified pipeline also provides significant improvements in the redshift measurements over the standard \emph{Redrock} pipeline for the challenging-to-identify missed QSOs. This point is most clearly illustrated by focusing on the redshifts of the missed QSOs in the sparse VI: see Fig.~\ref{fig:zdiff_missed} (top) where we compare the modified and standard pipeline redshifts of the high-quality missed QSOs to the VI redshifts. Unlike the standard \emph{Redrock} pipeline, only a modest fraction of missed QSOs identified by the modified pipeline have catastrophic redshift failures, and QSOs are reliably identified out to $z\approx$~4. 

\begin{table}
\caption{Expected recovery rates of high-quality QSOs for the main DESI survey}
\label{table:recovery_high_qual_dark}
%\centerline{
\hskip-1.7cm
\begin{tabular}{c|cc|cc}
\hline
 & \multicolumn{2}{c}{Standard pipeline} & \multicolumn{2}{c}{Modified pipeline} \\
\hline
Selection &  VI QSOs & VI QSOs & VI QSOs & VI QSOs \\
 & recovered & missed & recovered & missed \\
\hline
All QSOs & 85.5$\pm$0.5\% & 14.5$\pm$0.5\% & 93.6$\pm$0.4\% & 6.4$\pm$0.4\% \\
\hline
\end{tabular}
%}
\tablecomments{The percentage of high-quality (VI~$>2.5$) QSOs recovered as QSOs (or alternatively not recovered; i.e.,\ missed) by either the standard \emph{Redrock} or modified pipeline for the main-target selection, calculated following Eqn.~3. The listed error corresponds to a 1~$\sigma$ uncertainty for a binomial distribution.}
\end{table}

\begin{table*}
\caption{Key metrics for the main DESI survey}
\label{table:redrock_results_high_qual_dark}
\centerline{
\hskip-2.5cm\begin{tabular}{c|cccc|cccc}
\hline
 & \multicolumn{4}{c}{Standard pipeline} & \multicolumn{4}{c}{Modified pipeline} \\
\hline
QSO selection & \multicolumn{2}{c}{Good redshift purity} & Redshift & Redshift & \multicolumn{2}{c}{Good redshift purity} & Redshift & Redshift\\
 & ($dz=0.0033$) & ($dz=0.010$) & precision & accuracy & ($dz=0.0033$) & ($dz=0.010$) & precision & accuracy\\
\hline
All QSOs & 94.0$\pm$0.4\% & 98.6$\pm$0.2\% & 99$\pm$2~{\rm km/s} & 70$\pm$20~{\rm km/s} & 93.3$\pm$0.4\% & 99.0$\pm$0.1\% & 94$\pm$2~{\rm km/s} & 64$\pm$21~{\rm km/s} \\
$z<2.1$ QSOs & 95.4$\pm$0.4\% & 98.6$\pm$0.2\% & 109$\pm$3~{\rm km/s} & 22$\pm$17~{\rm km/s} & 95.5$\pm$0.4\% & 99.3$\pm$0.1\% & 98$\pm$2~{\rm km/s} & 24$\pm$18~{\rm km/s} \\
Tracer QSOs & 96.0$\pm$0.4\% & 99.4$\pm$0.1\% & 124$\pm$3~{\rm km/s} & 24$\pm$19~{\rm km/s} & 95.2$\pm$0.4\% & 99.5$\pm$0.1\% & 121$\pm$3~{\rm km/s} & 28$\pm$20~{\rm km/s} \\
Ly~$\alpha$ QSOs & 90.5$\pm$0.8\% & 98.7$\pm$0.3\% & 69$\pm$4~{\rm km/s} & 344$\pm$47~{\rm km/s} & 87.8$\pm$0.9\% & 98.2$\pm$0.4\% & 78$\pm$5~{\rm km/s} & 342$\pm$53~{\rm km/s} \\
\hline
\end{tabular}}
\tablecomments{Key metrics calculated for the main DESI survey using the standard \emph{Redrock} pipeline and modified pipeline for all QSOs, $z<2.1$ QSOs, tracer QSOs ($z=$~0.9--2.1), and Ly~$\alpha$ QSOs ($z>2.1$). The good redshift purity is calculated following Eqn.~4 for two different $dz$ thresholds; the listed error corresponds to a 1~$\sigma$ uncertainty for a binomial distribution. The redshift precision is calculated following Eqn.~6 and quantifies the 1~$\sigma$ variation in velocity dispersion between redshift pairs; the listed error is calculated from bootstrapping the sample 500 times. The redshift accuracy is defined as the median DESI--SDSS redshift offset for high-quality DESI QSOs within the SDSS DR16 QSO catalog; the listed uncertainty is the standard error of the median.}
\end{table*}

\begin{figure*}
\center
\includegraphics[width=0.49\textwidth]{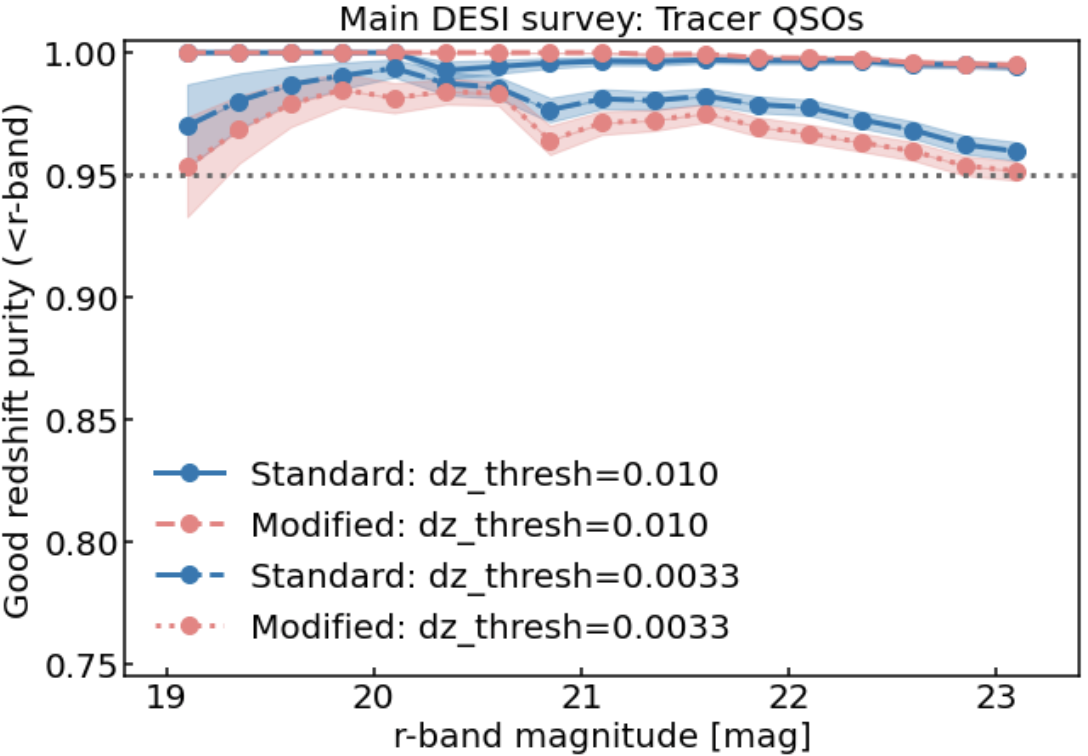}
\includegraphics[width=0.49\textwidth]{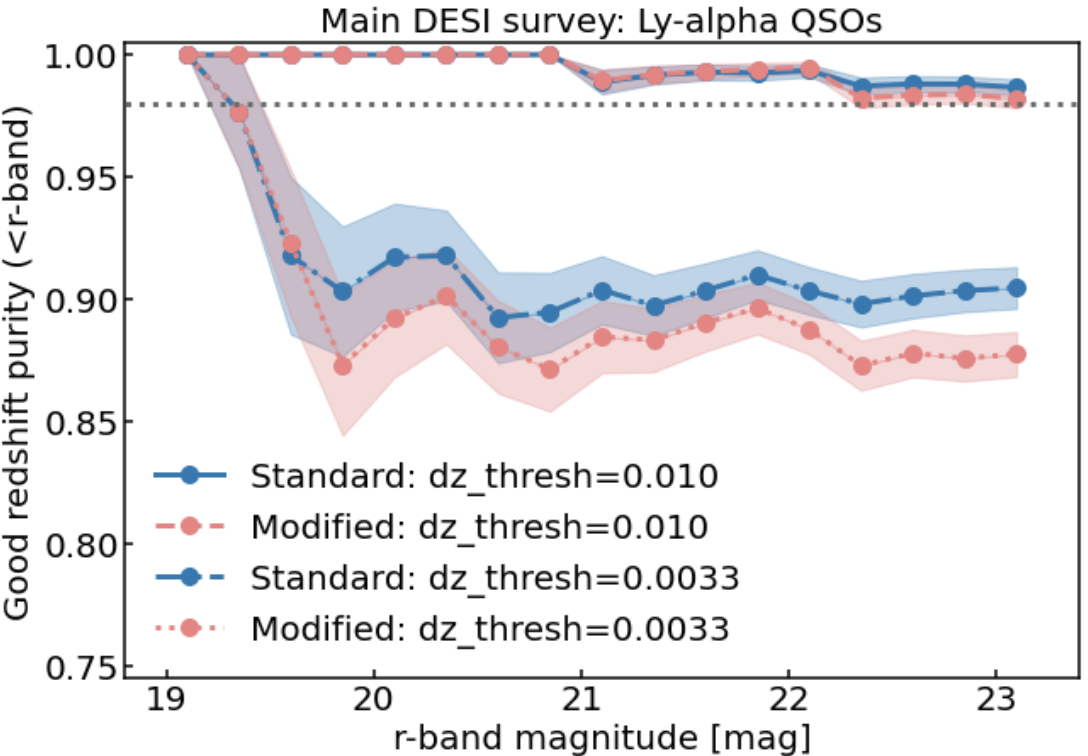}
\caption{Good-redshift purity versus $r$-band magnitude for main-selection targets identified as QSOs by either the standard \emph{Redrock} pipeline (blue) or the modified pipeline (red) for two $dz$ thresholds in dark-time exposures of $\approx$~1000~s for (left) tracer QSOs and (right) Ly~$\alpha$ QSOs. The cumulative fractions are plotted for dz$\_$thresh~=~0.0033 and dz$\_$thresh~=~0.010; the shaded regions indicate the 1~$\sigma$ uncertainty for a binomial distribution. The dotted horizontal line indicates the DESI scientific requirement.}
\label{fig:purity}
\end{figure*}

\section{Quantifying the performance of the main DESI survey with the visual inspection data}

In this section we exploit the VI data to quantify the expected performance of the main 5~year DESI survey using $\approx$~1000~s spectra. Our tests of the pipelines and comparisons of the SV1 versus main target selections in \S3 were based on the deep SV spectra with exposure times up-to an order of magnitude longer than the main-survey spectra. As a consequence, the results presented so far will be more optimistic than those achievable at the shorter main-survey exposures. However, since each SV spectrum was produced by coadding sets of shorter-exposure spectra, we can use these data to construct spectra with exposure times equivalent to those of the main survey. For our analyses here we constructed spectra with dark-time exposure times of 800--1200~s ($\approx$~1000~s) for the deep-field targets meeting the main-target selection criteria. 

Our basic approach and analyses follow those adopted for the galaxy surveys (Lan et~al.\ 2022) with the key difference that we only focus on the spectroscopically identified QSOs rather than using all targets. We primarily take this approach due to the excellent reliability with which our spectroscopic pipelines identify high-quality QSOs; see Table~\ref{table:redrock_results_deep}. However, the lower redshifts of the galaxies, which overlap with the more reliable galaxy samples from the BGS, LRG, and ELG surveys, also makes them less valuable cosmological tracers than the QSOs. For our analyses we define two QSO subsamples, in addition to the full QSO sample: ``tracer QSOs" at $z=$~0.9--2.1 and ``Ly~$\alpha$ QSOs" at $z>2.1$. 

The majority of our analyses are focused on characterizing the reliability of the spectroscopic redshifts for QSOs identified from both the standard or modified pipelines: in \S4.2 we measure the good-redshift purity, in \S4.3 we calculate the redshift precision, and in \S4.4 we measure the overall redshift accuracy. However, in \S4.1 we first calculate the overall recovery rate of QSOs to provide a benchmark against the QSO recovery rate using the full-depth spectra. We compare our results against the DESI scientific requirements (Abareshi et~al.\ 2022; DESI collaboration et~al.\ 2022).

\subsection{QSO recovery rate}

The QSO recovery rate effectively quantifies the ``completeness" of the QSO identification against the visually inspected QSO sample. We calculate the QSO recovery rate as

\begin{equation}
    {\rm QSO \ recovery \ rate=\frac{N_{\rm RR:VI,QSO}(VI\ \ge 2.5)}{N_{\rm VI,QSO}(VI\ \ge 2.5)},}
    \label{eq:recovery}
\end{equation}

\noindent where RR:VI refers to \emph{Redrock}-identified QSOs that are spectrally confirmed as QSOs from the VI. The results are presented in Table~\ref{table:recovery_high_qual_dark}. The modified pipeline recovers $\approx$~94\% of the visually inspected QSOs, a significant improvement over the standard \emph{Redrock} pipeline ($\approx$~86\% recovered). As expected, these results are worse than those achieved using the coadded spectra (modified pipeline: $\approx$~99\% recovered; standard pipeline: $\approx$~89\% recovered); see Table~\ref{table:redrock_results_deep}.

\begin{figure*}
\center
\includegraphics[width=0.49\textwidth]{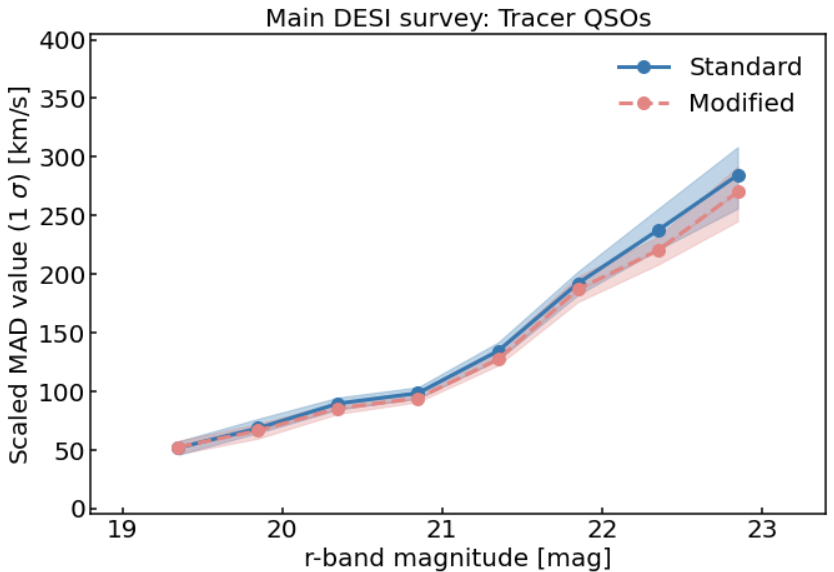}
\includegraphics[width=0.49\textwidth]{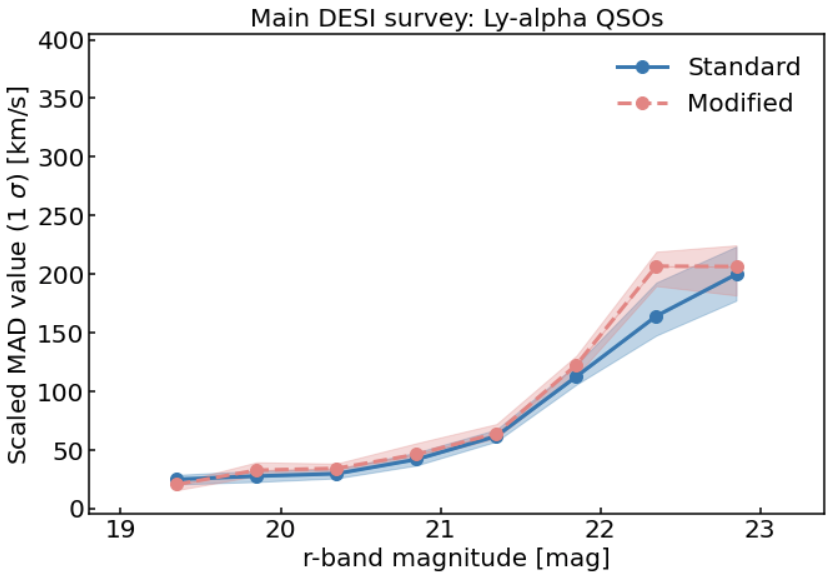}
\caption{Median absolute deviation of the velocity offsets ($\Delta{v}$) between redshift pairs versus $r$-band magnitude for main-selection targets identified as QSOs by either the standard \emph{Redrock} pipeline (blue) or the modified pipeline (red) in dark-time exposures of $\approx$~1000~s for (left) tracer QSOs and (right) Ly~$\alpha$ QSOs. The shaded regions indicate the 1~$\sigma$ uncertainty calculated via bootstrapping the sample 500 times.}
\label{fig:MAD}
\end{figure*}

\subsection{Good-redshift purity}

A key DESI metric is the fraction of QSO spectra with high-quality (VI~$\ge2.5$) redshifts within a given $dz$ threshold (dz$\_$thresh). We refer to this quantity as the ``good redshift purity", which we define as

\begin{align}
    & {\rm Good\ redshift\ purity=}\\ \nonumber 
    & {\rm \frac{N_{\rm RR,QSO}(VI\ \ge 2.5\ \&\ {dz_{\rm VI,RR}}\ \le \ {dz\_thresh})}{N_{\rm RR,QSO}}}
    \label{eq:purity}
\end{align}

\noindent where $dz_{\rm VI,RR}$ is the relative VI--RR redshift offset calculated following Eqn.~1 and replacing $z_{\rm A}$ and $z_{\rm B}$ with  $z_{\rm VI}$ and $z_{\rm RR}$, respectively. In our companion galaxy VI paper a threshold of dz$\_$thresh~$=$~0.0033 (equivalent to $dv=1000$~km~s$^{-1}$) was used (Lan et~al.\ 2022). However, given the greater uncertainty in measuring redshifts from broad QSO emission lines we also consider the more liberal threshold of dz$\_$thresh~$=$~0.010 (equivalent to $dv=3000$~km~s$^{-1}$). Our good-redshift purity results are shown in Table~\ref{table:redrock_results_high_qual_dark}.

The DESI scientific requirements for this analysis is framed as the ``catastrophic redshift failure" which is related to Eqn.~4 as:

\begin{align}
    & {\rm Catastrophic\ redshift\ failure=}\\ \nonumber 
    & 1 - {\rm Good\ redshift\ purity}
    \label{eq:catastrophic}
\end{align}

The required catastrophic redshift failure rate for the tracer QSOs is $<5$\% (good redshift purity of $>95$\%) for dz$\_$thresh~=~0.0033, which is achieved for both pipelines; see Table~\ref{table:redrock_results_high_qual_dark}. The required catastrophic redshift failure rate for the Ly~$\alpha$ QSOs is $<2$\% (good redshift purity of $>98$\%) which is also achieved for both pipelines with dz$\_$thresh~=~0.010.

These analyses have quantified the overall good-redshift purity but we would also expect these results to vary with data quality. In our companion galaxy paper (Lan et~al.\ 2022), the good-redshift purity was calculated as a function of $\Delta\chi^2$, which is defined as the difference in $\chi^2$ between the best fitting and second-best fitting \emph{Redrock} template-redshift solution for each target: high $\Delta\chi^2$ values are more likely to correspond to higher-quality spectra. We do not consider $\Delta\chi^2$ in our analyses here partly because \emph{Redrock} identifies high-quality QSOs down to the $\Delta\chi^2=0$ limit with just a small fraction of low-quality spectra. However, adopting this approach would also limit our analyses to just the standard \emph{Redrock} pipeline since the modified pipeline takes a more complex approach in the calculation of redshifts, adopting a tight redshift prior using the QN redshift but only for a fraction of the QSOs. Instead, to provide insight on the dependence of the spectroscopic redshift accuracy with data quality, we use the $r$-band magnitude as a ``proxy" of the data quality.

In Fig.~\ref{fig:purity} we plot the good-redshift purity as a function of $r$-band magnitude for both the tracer and Ly~$\alpha$ QSOs to show how the results depend on both data quality and dz$\_$thresh. The required good-redshift purity of $>95$\% is achieved for the tracer QSOs over the full $r$-band magnitude range for both $dz$ thresholds with both pipelines.\footnote{We note that the ``choppy" behaviour of the tracks is due to the small number of high-quality VI'ed QSOs, particularly at bright magnitudes.} A broadly similar trend is also seen for the Ly~$\alpha$ QSOs, although the good-redshift purity varies more greatly with $dz$ than for the tracer QSOs, from $<90$\% for dz$\_$thresh~=~0.0033 to $>96\%$ for dz$\_$thresh~=~0.010 regardless of the pipeline. The good-redshift purity of the tracer QSOs decreases with $r$-band magnitude, as expected if data quality drives the overall redshift uncertainties. A similar behaviour is also seen for the Ly~$\alpha$ QSOs except for dz$\_$thresh~=~0.0033 where the good-redshift purity is comparatively flat at $\approx$~85--90\% over almost the full plotted $r$-band magnitude range. This shows that \emph{Redrock} is unable to reliably measure redshifts within $dz<0.0033$ for $\approx$~10--15\% of the Ly~$\alpha$ QSOs, unrelated to data quality which is likely due to complexity in the rest-frame UV spectra for a subset of the QSOs (e.g.,\ complex and poorly defined emission peaks due to broad emission and absorption features).

\begin{figure}
\center
\includegraphics[width=0.47\textwidth]{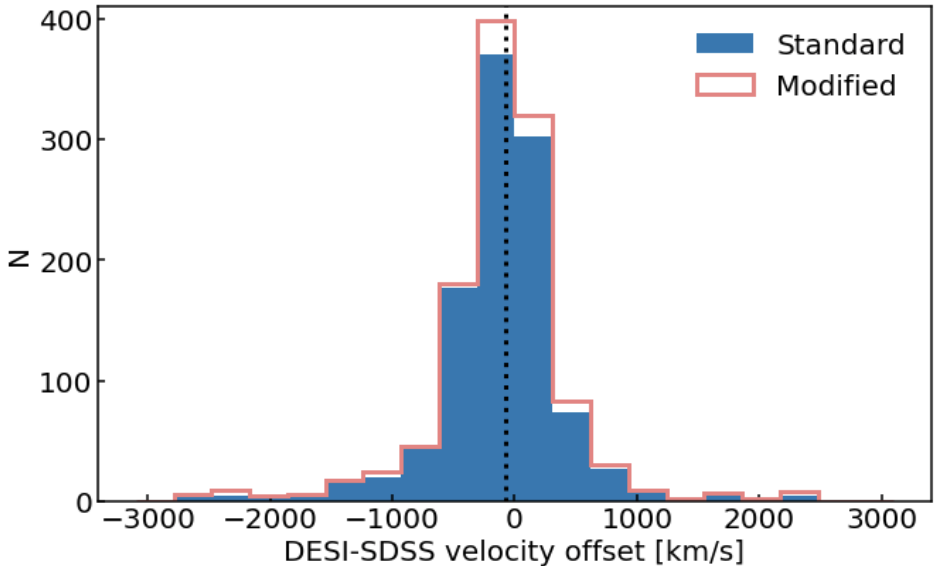}
\caption{Assessment of redshift accuracy (expressed as the velocity offset, $dv$, in km~s$^{-1}$) for main-selection targets with dark-time exposures of $\approx$~1000~s matched to the SDSS DR16 quasar catalog for the standard \emph{Redrock} pipeline (blue) and the modified pipeline (red). The dotted vertical line indicates the median velocity offset.}
\label{fig:dz_sdss}
\end{figure}

\subsection{Redshift precision}

The redshift precision quantifies the random error in the QSO redshift measurements. The DESI scientific requirement for this parameter is for the random error between individual QSO redshifts to be $\Delta{z}<$~0.0025(1~+~$z$), equivalent to velocity offsets of $\Delta{v}<750$~km~s$^{-1}$, within a Gaussian core (i.e.,\ 1~$\sigma$). In this calculation we exploit the multiple short-exposure redshift measurements available for each target to calculate the redshift offset (expressed as the velocity offset, $dv$, in km~s$^{-1}$) between each redshift pair (${i,j}$)

\begin{equation}
    dv_{i,j}=\frac{z_{\rm RR,i}-z_{\rm RR,j}}{(1+z_{\rm RR,i})}\times{c}.
    \label{eq:dv}
\end{equation}

\noindent The random error is then estimated as the dispersion from the distribution of redshift-pair offsets. Following Lan et~al.\ (2022), we calculate the median absolute deviation (MAD), rather than the standard deviation, to minimize the impact of outlier measurements. We then scale the MAD value by 1.4828 to represent the 1~$\sigma$ error and divide by $\sqrt{2}$ to take account of the fact that the dispersion includes measurement errors from two redshifts:

\begin{equation}
    {\rm Scaled\ MAD\ value}=\frac{{\rm MAD}\times\ 1.4828}{{\sqrt{2}}}.
    \label{eq:dv}
\end{equation}

The results are shown in Table~\ref{table:redrock_results_high_qual_dark} and the variation in the scaled MAD value with $r$-band magnitude is presented in Fig.~\ref{fig:MAD}. Overall the DESI scientific requirements are met for all QSO subsets with scaled MAD values of $\Delta{v}<$~300~km~s$^{-1}$ for even the faintest QSOs; however, clear trends with $r$-band magnitude are seen for both the tracer and Ly~$\alpha$ QSOs, demonstrating the general impact of data quality on the precision of individual redshift measurements. No significant differences are seen between the two pipelines.

\subsection{Redshift accuracy}

The redshift accuracy quantifies the systematic accuracy of the QSO redshifts. The DESI scientific requirement is for a redshift accuracy of $\Delta{z}<$~0.0004(1~+~$z$), equivalent to velocity offsets of $\Delta{v}<120$~km~s$^{-1}$, in the redshifts of the tracer QSOs. To provide an assessment of the redshift accuracy we matched the DESI QSOs to the SDSS DR16 quasar catalog (Lyke et~al.\ 2020) using a 5$^{\prime\prime}$ search radius and calculated the median redshift offset. Overall, we found matches to 292 SDSS quasars of which 176 are tracer QSOs (a further 84 are Ly~$\alpha$ QSOs). In our analyses we used the $\approx$~1000~s DESI spectra for these matched quasars which increases the source statistics by a factor $\approx$~5.

The overall measured median velocity offsets for all matched QSOs are $\approx$~70~km~s$^{-1}$ for both pipelines; see Table~\ref{table:redrock_results_high_qual_dark} and Fig.~\ref{fig:dz_sdss}. The median velocity offsets for the tracer QSOs are even smaller at $\approx$~30~km~s$^{-1}$, meeting the DESI scientific requirements for both pipelines and consistent with the standard error of the median. As shown in Table~\ref{table:redrock_results_high_qual_dark}, the larger velocity offset measurements for the all-matched QSO sample is due to the Ly~$\alpha$ QSOs, which have median velocity offsets of $\approx$~340~km~s$^{-1}$. This same effect is remarked upon in the DESI QSO target-selection paper (see \S7.3 of Chaussidon et~al. 2022) and the cause is investigated extensively in the SDSS eBOSS BAO paper (see Appendix B of du Mas des Bourboux et~al. 2020). Two factors likely drive the systematic redshift inaccuracy of the Ly~$\alpha$ QSOs (1) Ly~$\alpha$ absorption features entering the DESI band pass and distorting the profile of the Ly~$\alpha$ emission line, and (2) Mg~II emission leaving the DESI band pass and, consequently, causing the pipeline redshift measurements to be more reliant on the less-reliable C~IV emission line (e.g.,\ see Tytler \& Fan 1992; Hewett \& Wild 2010). We note that since the SDSS DR16 quasar catalog is limited to $r<22$, the redshift accuracy may be worse for the faintest DESI targets.

\begin{figure*}
\center
\includegraphics[width=0.49\textwidth]{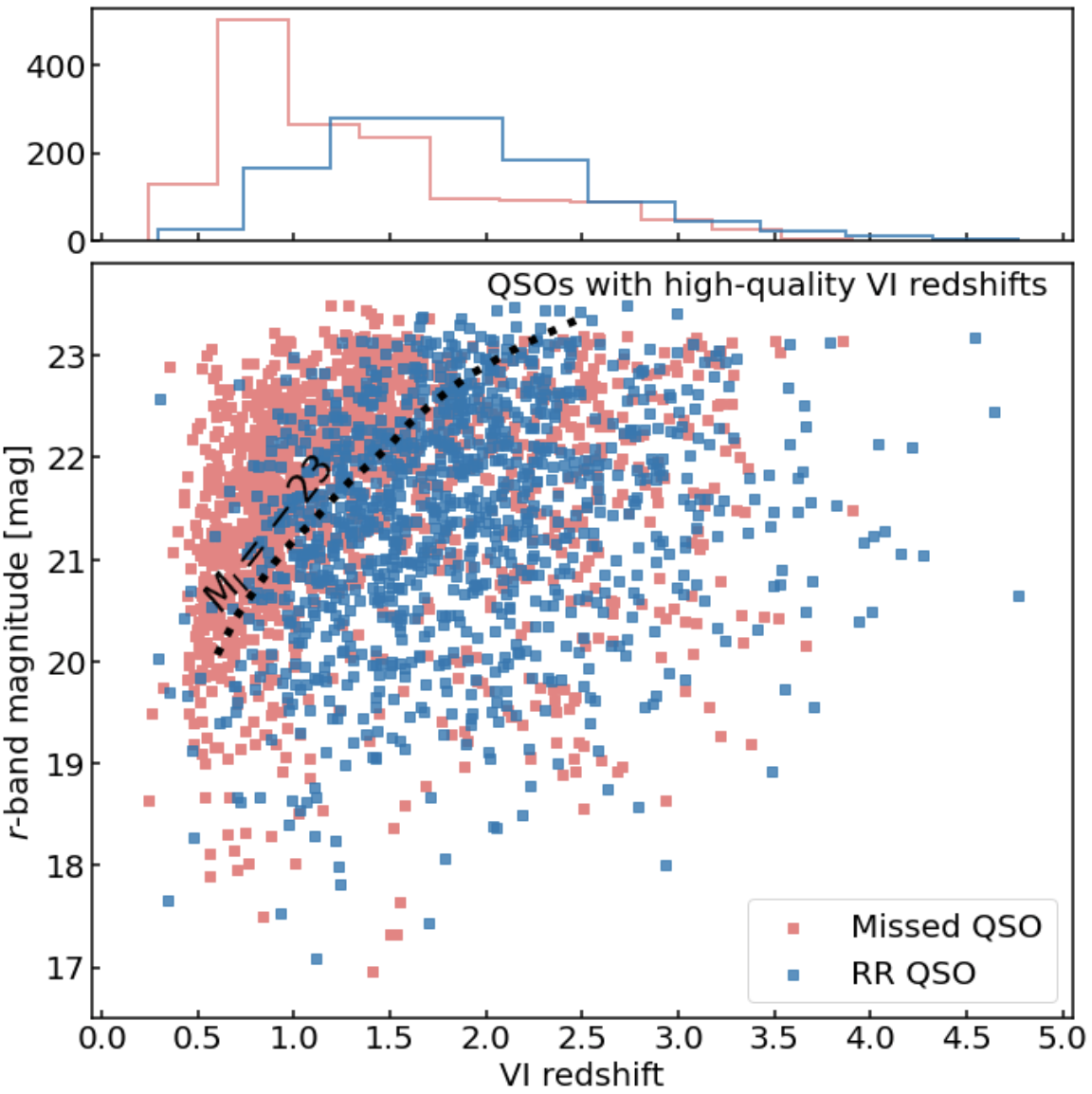}
\includegraphics[width=0.49\textwidth]{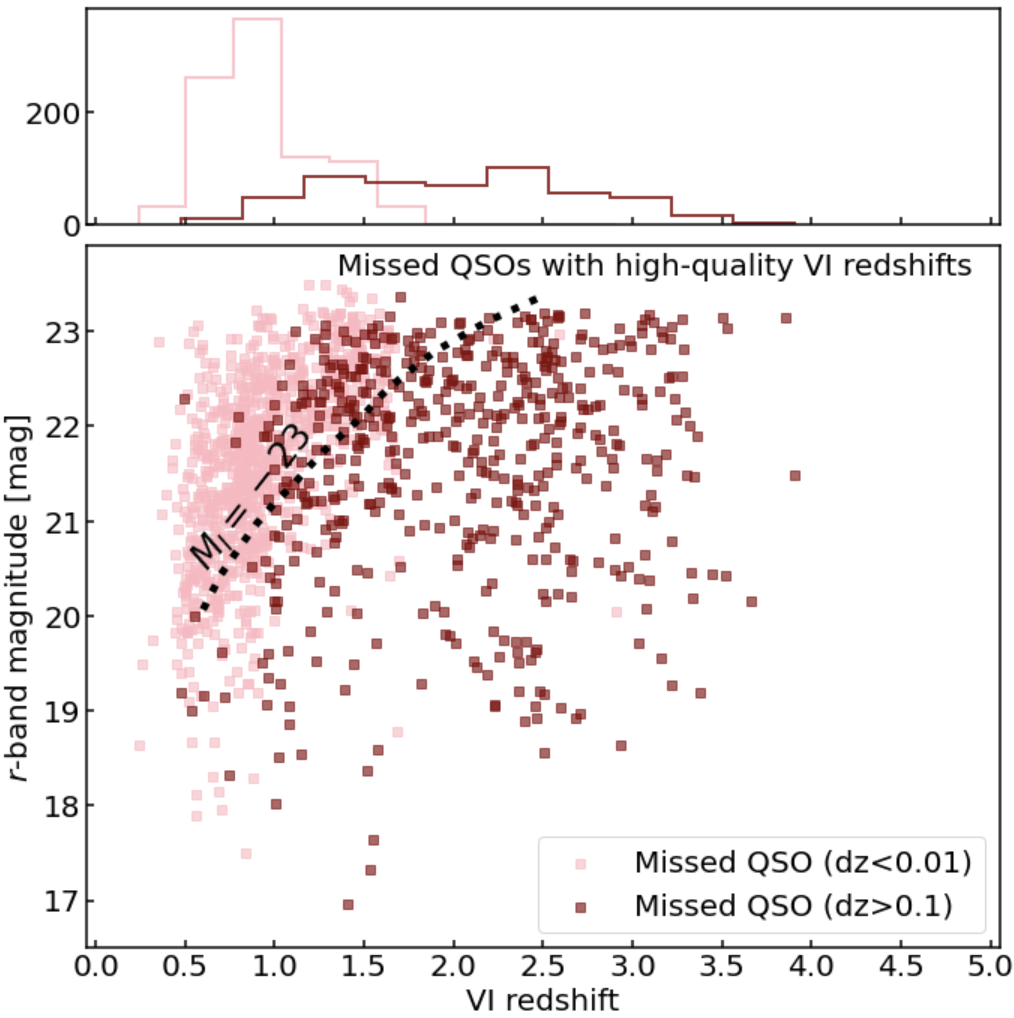}
\caption{$r$-band magnitude versus redshift for (left) high-quality \emph{Redrock} identified QSOs from the deep field VI (blue squares) and the missed QSOs in the sparse-field VI (red squares) and (right) high-quality missed QSOs from the sparse VI, split between those with reliable redshifts ($dz<0.01$: pink squares) and those with large catastrophic redshift failures ($dz>0.1$: maroon squares); we note that missed QSOs with $dz=0.01-0.1$ are not plotted in the right-hand panel. The black dashed curve shows the expected track for a $M_{\rm I}=-23$~mag QSO to provide a basic discrimination between classical QSOs and lower luminosity Seyfert-type galaxies. The top panels show the redshift distributions for the (left) high-quality \emph{Redrock} QSOs (blue) and missed QSOs (red) and (right) high-quality missed QSOs with reliable redshifts (pink) and redshift failures (maroon).}
\label{fig:magzdouble}
\end{figure*}

\begin{figure*}
\center
\includegraphics[width=0.75\textwidth]{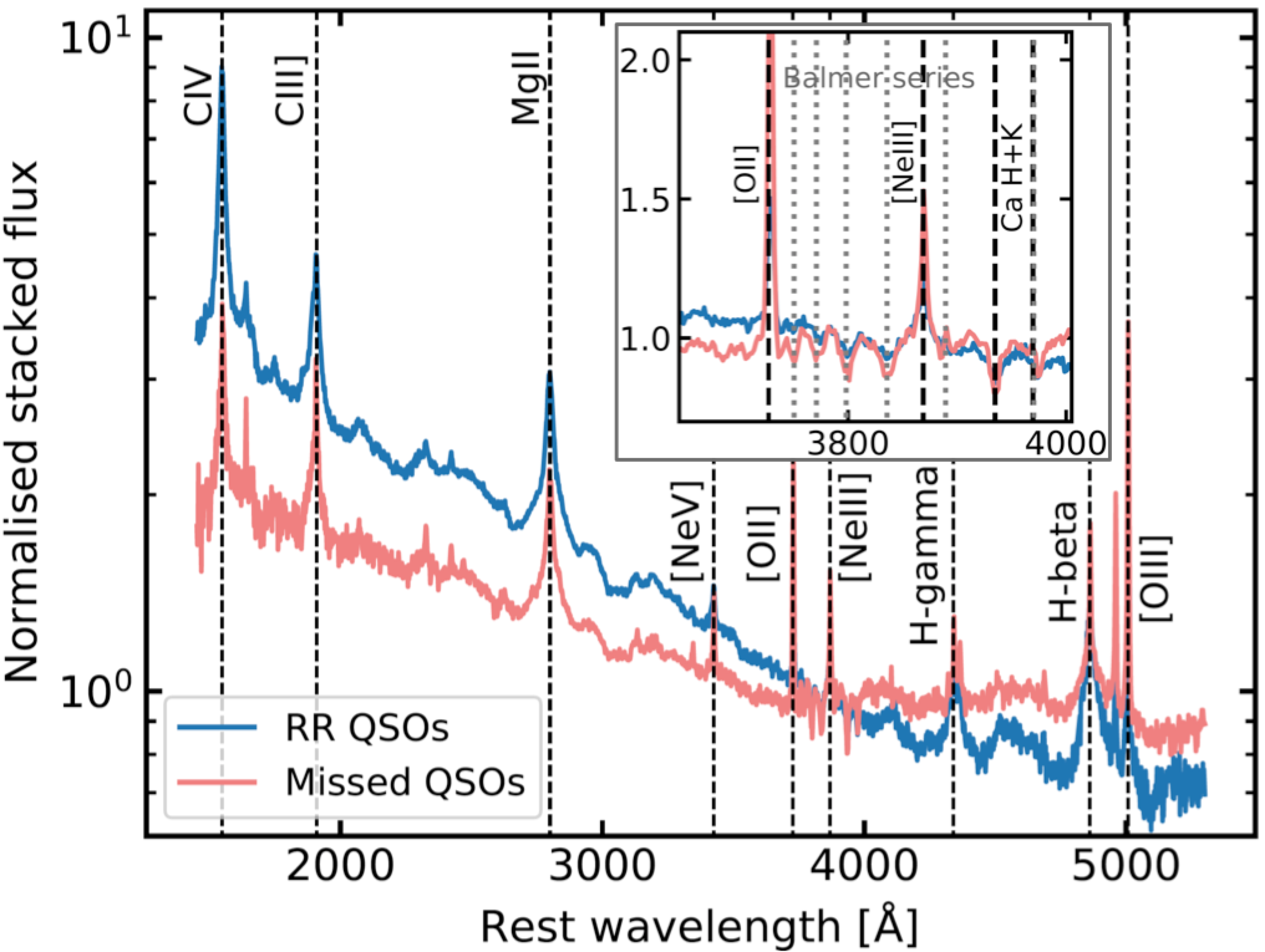}
\caption{Median composites of the \emph{Redrock}-identified QSOs from the deep-field VI (blue) and QSOs missed by \emph{Redrock} but identified from the afterburner approaches in the sparse-field VI (red). The composite spectra are normalized over rest-frame 3800--3900\AA\ for visualization purposes. The inset plot is focused around 3700--4000\AA\ to highlight the prominent stellar absorption features in the missed QSO sample. The vertical lines indicate several key emission and absorption features.}
\label{fig:composites}
\end{figure*}

\begin{figure*}
\center
\includegraphics[width=1.0\textwidth]{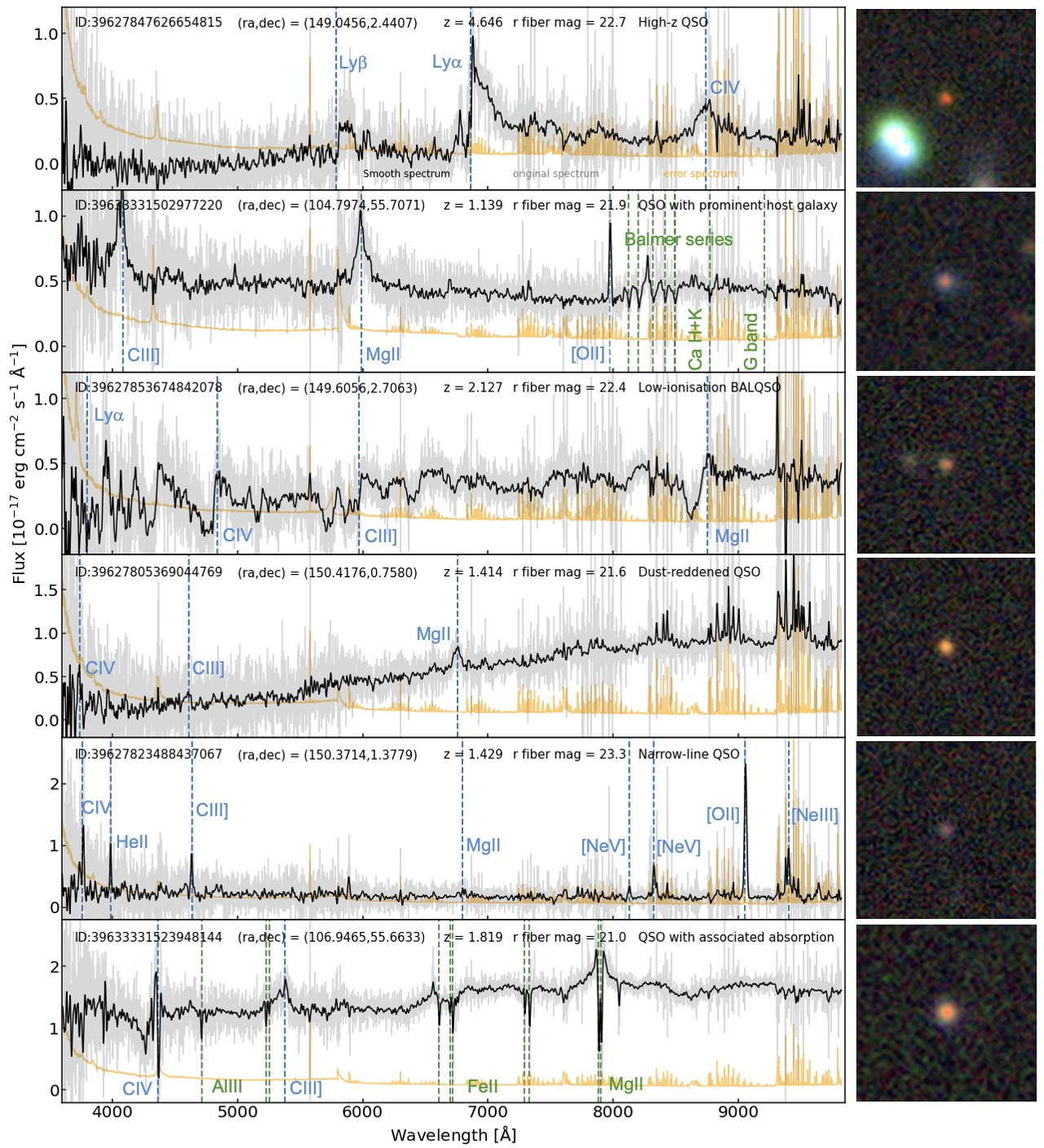}
\caption{Compilation of DESI spectra to illustrate some of the diversity in systems identified in the QSO survey (left) and thumbnail images ($18^{\prime\prime}\times18^{\prime\prime}$) centered on each target (right). Salient information for each target is plotted at the top of each spectrum including a description of the QSO sub type. For each target both an unsmoothed (grey) and smoothed (black) spectrum are plotted along with the associated error spectrum (orange). Some of the most prominent emission lines (blue vertical dashed lines) and absorption lines (green vertical dashed lines) are highlighted.}
\label{fig:examples_diversity}
\end{figure*}

\section{Spectral diversity in the DESI quasar survey}

As is clear from the VI results presented in \S3, one of the challenges in the QSO survey is the identification of QSOs from stars and the, sometimes more dominant, galaxy populations. In this section we explore some of the spectral diversity in the DESI QSO survey. In \S5.1 we investigate how the QSOs identified by \emph{Redrock} differ from the missed QSOs identified by the afterburners to understand why some QSOs are misidentified by \emph{Redrock}. In \S5.2 we investigate the optical spectra in more detail to demonstrate the spectral diversity of the visually identified QSOs by selecting interesting example spectra. In \S5.3 we focus on the high-quality galaxies and investigate how they differ from the high-quality QSOs and whether a significant number of QSOs remain unidentified. Finally, in \S5.4 we briefly investigate the incidence of two sources contributing to the same DESI spectrum and show several example spectra from the VI.

\subsection{Investigating the differences between the Redrock quasars and the missed quasars}

In Fig.~\ref{fig:magzdouble} (left) we plot the $r$-band magnitude versus redshift for the QSOs identified by \emph{Redrock} in the deep-field VI and compare them to the afterburner-identified missed QSOs from the sparse VI. The missed QSOs have a more strongly peaked redshift distribution than the \emph{Redrock} QSOs due to the enhanced ``clump" of systems in the $r$-band--redshift plane, also seen in Fig.~\ref{fig:r-band} (right). The missed QSOs also fill out a fainter region in the $r$-band magnitude--redshift plane in Fig.~\ref{fig:magzdouble} (left) than the \emph{Redrock} QSOs: at any given redshift out to $z\approx$~1.7, the missed QSOs are identified to fainter $r$-band magnitudes than the \emph{Redrock}-identified QSOs. Consequently, the inclusion of the afterburners leads to a much more comprehensive QSO census, particularly at $z<1.7$ due to the identification of optically fainter systems. 

In Fig.~\ref{fig:magzdouble} (left) we also plot the magnitude--redshift tracks for a $M_{\rm I}=-23$~mag QSO to provide a basic descrimination between luminous QSOs and lower-luminosity systems. As can be seen, many of the $z<1.7$ missed QSOs lie above the $M_{\rm I}=-23$~mag QSO tracks indicating that they are probably relatively low-luminosity QSOs, more analogous to Seyfert galaxies identified in the local Universe than classical QSOs. Indeed, for the majority of the $z<1.7$ missed QSOs, \emph{Redrock} identifies the correct redshift but classifies the target as a galaxy, consistent with that expected for a low-luminosity QSO. We visually demonstrate this in Fig.~\ref{fig:magzdouble} (right) where we again plot $r$-band magnitude versus redshift but now split the missed QSOs on the basis of $dz$: a clear division in the redshift distribution is seen with the missed QSOs at $z<1$ having reliable redshifts ($dz<0.01$) while the missed QSOs at $z>1.7$ have predominantly large catastrophic redshift failures ($dz>0.1$), with a mix of reliable redshifts and redshift failures over $z\approx$~1.0--1.7. The missed QSOs with redshift failures have a broadly similar redshift distribution to the \emph{Redrock} QSOs (see Fig.~\ref{fig:magzdouble}) when the correct redshift is identified.

To further shed light on the physical origins of the differences between the \emph{Redrock} QSOs and the missed QSOs, we constructed median composite spectra for both samples by stacking the rest-frame optical spectra; see Fig.~\ref{fig:composites}.\footnote{Each composite spectrum is constructed following the approach outlined in Fawcett et~al.\ (2022). Briefly, the ends of each contributing spectrum are trimmed to remove noisy data and then corrected for Galactic extinction and shifted to rest-frame wavelengths using the VI redshift. Each spectrum is then adjusted to a common wavelength grid and normalised at rest-frame 3000~\AA\ across a 20~\AA\ window; the normalization wavelength was chosen to maximise the number of sources with observed-frame coverage corresponding to rest-frame 3000~\AA. The composite is then created by taking the median across all spectra contributing to a given wavelength bin, applying a minimum threshold of 30 spectra/bin.} The \emph{Redrock} QSO composite has all of the features of a classical QSO: prominent broad emission lines and a strongly rising continuum to rest-frame UV wavelengths. However, by comparison, the missed QSO composite has a much flatter continuum slope with a deficit of emission at shorter wavelengths, enhanced emission at longer wavelengths, and stronger narrow forbidden lines (e.g.,\ [Ne~V] and [O~II]). The frequency dependent drop in the continuum emission to UV wavelengths for the missed QSOs is the characteristic signature of dust extinction along the line of sight. To provide insight on the origin of the enhanced longer-wavelength emission we produced a zoom in of the spectral region around $\approx$~3800~\AA; see Fig.~\ref{fig:composites} (inset). The missed QSOs have stronger Balmer and Ca H+K absorption lines, the expected signatures from a significant host-galaxy contribution, which also explains the enhanced continuum longward of 4000~\AA\ (i.e.,\ the emission from stars in the host galaxy). 

Overall, the reason the missed QSOs were misidentified by the standard \emph{Redrock} pipeline appears to be due to the significant contribution from the host galaxy and/or the suppression of the rest-frame UV emission from dust extinction. The requirement for \emph{Redrock} to fit the DESI spectra across all of the galaxy surveys, in addition to the QSO survey, limits the range of QSOs templates that can be effectively utilized and, consequently, it is not possible to account for the full diversity of the QSO spectral class; see also Footnote~6. \emph{Redrock} can therefore measure the correct redshift of a red QSO when prominent host-galaxy features are present in the optical spectrum, as is the case for low-luminosity QSOs up-to $z\approx1.7$ (i.e.,\ distant analogs of Seyfert galaxies), where the majority of the strongest host-galaxy features lie within the DESI spectral bandpass. Therefore, it is unsurprising that the vast majority of the missed QSOs at $z<1$, and a significant fraction of missed QSOs out to $z\approx$~1.7, have reliable redshifts even though the standard \emph{Redrock} pipeline identifies them as galaxies instead of QSOs; the often noisy data at $>8500$~\AA\ (see Fig.~\ref{fig:example_spectra} for examples) is likely responsible for the decrease in the reliable redshift fraction for missed QSOs at $z\approx$~1.0--1.7. At higher redshifts, the strongest host-galaxy features move out of the DESI spectral bandpass and, without a red QSO template solution, \emph{Redrock} cannot reliably measure the target redshift.

\subsection{A glimpse of the spectral diversity in the QSO survey}

The composite spectra shown in Fig.~\ref{fig:composites} demonstrate the broad diversity between the QSOs identified by the standard \emph{Redrock} pipeline and the missed QSOs recovered by the afterburners. However, they do not reveal the full range of spectral diversity within the QSO survey. To provide a glimpse of the overall spectral diversity of targets in the QSO survey, in Fig.~\ref{fig:examples_diversity} we plot the DESI spectra for individual QSOs, selected to cover a broad range in spectral diversity.

The first spectrum is an example of an optically faint high-redshift QSO. Many high-redshift QSOs have been identified in previous QSO surveys; however, the fainter optical magnitude limit of DESI allows for the reliable identification of lower luminosity, and therefore more typical, systems. The second spectrum is an example of a QSO with a prominent host galaxy, broadly similar to the missed QSO composite shown in Fig.~\ref{fig:composites}: the strong host-galaxy absorption features are highlighted in green and are consistent with a post starburst. The identification of the host galaxy is due to both the faint optical magnitude limit of DESI and the relatively high spectral resolution, which provide the potential to characterize the host-galaxy properties (e.g.,\ luminosity weighted stellar age and host-galaxy mass) for a significant fraction of the DESI QSOs. The third spectrum is a spectacular example of an optically faint low-ionisation broad-absorption line QSO (BALQSO), where strong broad absorption troughs are seen blueward of both the low-ionisation (Mg~II) and high-ionisation (C~IV; C~III]) broad emission lines. DESI is able to identify BALQSOs down to fainter optical magnitudes than previous optical QSO surveys. On the basis of the VI, $>10$\% of the high-quality deep-field QSOs showed visual evidence for broad-absorption line features at $z>1.57$. The fourth spectrum is an example of an individual dust-reddened QSO missed by the standard \emph{Redrock} pipeline but identified by the afterburners. The faint optical magnitude of DESI combined with the optical--mid-IR colour selection allows for the identification of redder and fainter QSOs than those identified in the SDSS. The fifth spectrum shows a QSO where only narrow lines are detected: the identification of the strong high-excitation [Ne~V] emission line provides the evidence that this is a QSO as opposed to a galaxy. Due to the requirement for a point-source optical morphology, the QSO survey will only identify a subset of the narrow-line QSO population since the majority are expected to have extended optical morphologies (i.e.,\ due to optical emission being dominated by the host galaxy). The last spectrum shows a QSO with strong associated absorption features: in this example, a significant fraction of the absorption appears to be due to the host-galaxy environment since absorption features are identified at the systemic redshift of the QSO. The relatively high spectral resolution of DESI allows for the comprehensive identification of narrow absorption features. A comprehensive evaluation of Mg~II absorption systems identified in the QSO survey will be provided in Napolitano et~al.\ (in prep.).

\subsection{Investigating the nature of the galaxies detected in the QSO survey}

\begin{figure*}[t]
\center
\includegraphics[width=0.75\textwidth]{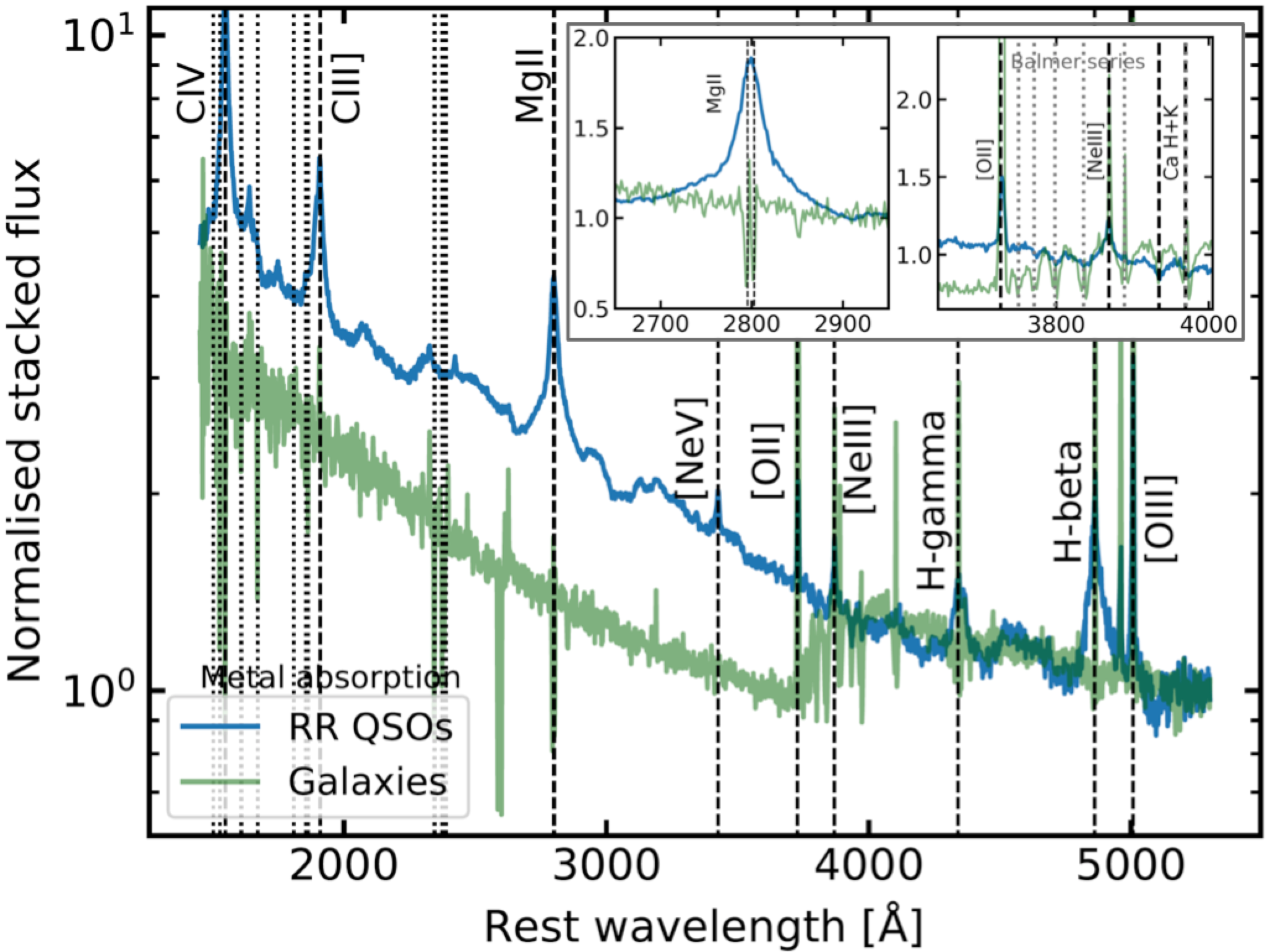}
\caption{Median composites of high-quality \emph{Redrock}-identified QSOs (blue) and visually confirmed galaxies (green) from the deep-field VI of the QSO survey. These composite spectra are normalized over rest-frame 5100--5200~\AA\ for visualization purposes. The inset plot is focused around 2600--3000~\AA\ and 3600--4000~\AA\ to highlight the lack of AGN features (left) and the prominent stellar absorption features (right) in comparison to the QSO composite. The vertical lines indicate several key emission (dashed line) and absorption (dotted line) features.}
\label{fig:composites_QSO_vs_gal}
\end{figure*}

The combination of the \emph{Redrock} and afterburner identified QSOs provides a comprehensive selection of QSOs. However, what is the nature of the galaxies detected in the QSO survey and could they host weak QSO or AGN activity? Following the same approach as for the QSO composites (see Footnote~8), we stacked the rest-frame spectra of the high-quality galaxies identified in the deep-field VI of the QSO survey to construct a median galaxy composite; see Fig.~\ref{fig:composites_QSO_vs_gal}. The galaxy composite shows a strong rise to UV wavelengths, similar to the \emph{Redrock} QSO composite, but lacks the associated QSO signature of broad emission lines. Furthermore, there is no evidence for [Ne~V] in the galaxy composite, a high-excitation emission line typically seen in AGN and QSOs; see Fig.~\ref{fig:composites_QSO_vs_gal} (inset). The galaxy composite is also distinguishable from the \emph{Redrock} QSO composite in having a strongly rising continuum to long wavelengths; the continuum rise is significantly stronger than that seen in the missed QSO composite. This continuum rise to long wavelengths is due to the stellar emission from the host galaxy, as is apparent from the very strong Balmer and Ca~H+K absorption features; see Fig.~\ref{fig:composites_QSO_vs_gal} (inset). Many absorption features at the wavelengths expected for metal lines (e.g.,\ Fe~II; Al~III; S~III) are also seen in the galaxy composite. 

What is the nature of these systems and what is the origin of the strong UV continuum, which is undoubtably the primary reason for their selection within the QSO survey? These systems are most likely compact galaxies undergoing significant star-formation activity, similar to the $z\approx$~0.6 post-starburst galaxies identified in the SDSS survey (Tremonti et~al.\ 2007). The galaxies explored in Tremonti et~al.\ (2007) had strong blue-shifted Mg~II absorption systems, indicative of powerful galactic winds with $v>500$~km~s$^{-1}$. Our galaxy composite shows Mg~II absorption but it is only slightly blue shifted. However, these DESI galaxies are several magnitudes fainter than those identified in the SDSS and, consequently, they likely host less-powerful galactic winds. Furthermore, the galaxy composite only provides an average Mg~II absorption constraint and some individual systems may show significantly stronger blue-shifted Mg~II absorption.

Overall, on the basis of this brief analysis, we do not find any clear evidence for additional significant QSO activity from the galaxy composite, despite the strong UV continuum emission. Weak QSO (or AGN) activity may be present in a small fraction of the systems, which would be missed by our analysis but could be revealed through more detailed inspection and spectral fitting of the individual DESI spectra. On the basis of the VI, AGN features (i.e.,\ [Ne~V]; strong [O~III]; C~III]) were noted by visual inspectors for just 8 ($\approx$~0.5\%) of the 1491 high-quality galaxies, although the true number of unidentified AGN could be much higher. Indeed, cross matching the 101 high-quality galaxies with sensitive {\it Chandra} observations in the COSMOS field (Marchesi et~al.\ 2016) revealed 5 X-ray detections with X-ray luminosities consistent with moderate--high luminosity AGN activity ($\approx10^{42}$--$10^{44}$~erg~s$^{-1}$; Brandt \& Alexander 2015), suggesting an X-ray AGN fraction within the galaxy population selected by the DESI quasar survey of $\approx$~5\%.

\subsection{Two sources contributing to a single spectrum}

\begin{figure*}
\center
\includegraphics[width=1.0\textwidth]{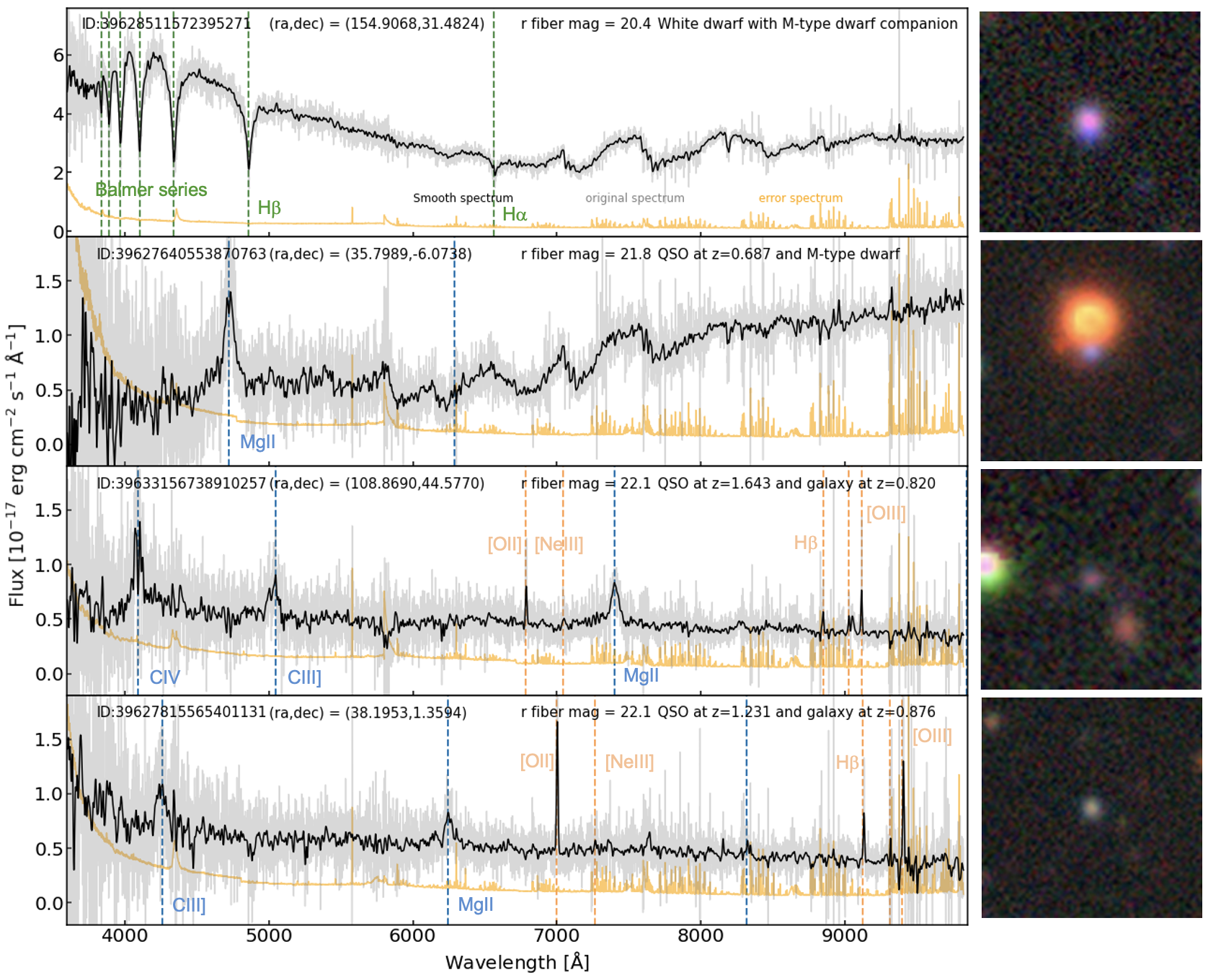}
\caption{Examples of systems where two objects are contributing to the DESI optical spectrum (left) and thumbnail images ($18^{\prime\prime}\times18^{\prime\prime}$) centered on each target (right). Salient information for each target is plotted at the top of each spectrum including a description of the two contributing objects. For each target both an unsmoothed (grey) and smoothed (black) spectrum are plotted along with the associated error spectrum (orange). All of the plotted spectra are high-quality with VI quality flags of 4. Some of the most prominent emission and absorption lines are highlighted using vertical dashed lines and plotted in green (stellar absorption features), blue (QSO emission lines), and orange (galaxy emission lines).}
\label{fig:examples_two_objects}
\end{figure*}

The vast majority of the optical spectra in DESI are produced by a single object. However, for a small fraction of the spectra, at least two objects contribute to a single optical spectrum. We show some examples in Fig.~\ref{fig:examples_two_objects} from the deep-field VI. The first spectrum shows a stellar binary system composed of a hot white dwarf and a cool M-type red dwarf; the overall continuum shape is similar to that of a QSO but the strong absorption features identify the two stars. The second spectrum shows a QSO spectrum that is contaminated at long wavelengths by an optically bright M-type red dwarf, which outshines the QSO in the associated finding chart. The other remaining spectra show the clear signatures of a higher-redshift QSO spectrum contaminated by a lower-redshift galaxy. Overall, on the basis of the VI, $\approx$~1\% of the optical spectra in the DESI quasar survey appear to be the superposition of at least two contributing targets. The systems where both targets are galaxies and/or QSOs can provide a rich dataset with which to probe the dark-matter component of the foreground target via strong lensing.

\section{Summary}

We have presented the first results from the visual inspection of the optical spectra obtained during the SV phase of the DESI QSO survey. The SV spectra are more sensitive than those obtained in the main 5~year survey (exposure times up-to an order of magnitude longer), allowing for the construction of reliable ``truth tables" with which to test the different target-selection approaches and spectroscopic pipelines. Furthermore, since the SV spectra are coadds of several shorter spectra, we also used the short-exposure spectra to characterize the reliability of the spectroscopic redshifts at the main 5~year survey depth. Our analyses have focused on (1) the complete VI of all targets within three deep tiles and (2) the sparse VI across 27 additional SV tiles of QSOs missed by \emph{Redrock} but identified by at least one of our three ``afterburner" approaches (Mg~II afterburner; SQUEzE; QuasarNet). Our main results are the following:

\begin{itemize}
    \item Overall the main target selection is much more efficient at selecting high-quality QSOs ($\approx$~71\%) than the SV1 target selection ($\approx$~34\%), although a non-negligible fraction of high-quality galaxies and stars are identified with both target-selection approaches (main: $\approx$~16\% galaxies; $\approx$~6\% stars; SV1: $\approx$~40\% galaxies, $\approx$~12\% stars). See \S3.1.
    \item The standard \emph{Redrock} pipeline reliably selects high-quality QSOs with a small fraction of low-quality contaminants but misses a non-negligible fraction ($>10\%$) of the visually identified QSOs. However, we can reliably recover the majority of these missed QSOs using ``afterburner" QSO-identification approaches. From the combination of \emph{Redrock} and the ``afterburners" we constructed a modified pipeline which is able to recover the majority of the missed QSOs while maintaining a small fraction of low-quality contaminants. See \S3.1--3.3.
    \item At the shallow depth of the main 5~year survey ($\approx$~1000~s) both the standard and modified pipelines exceed the DESI scientific requirements for good-redshift purity (assuming $dz<0.0033$), redshift precision, and redshift accuracy, with the exception of the Ly~$\alpha$ QSOs which achieves the required 98\% good-redshift purity for $dz<0.010$. However, the modified pipeline recovers a substantially larger fraction of the overall QSO sample ($\approx$~94\%) than the standard \emph{Redrock} pipeline ($\approx$~86\%). See \S4.
    \item The QSOs missed by the standard \emph{Redrock} pipeline have redder overall spectra than the \emph{Redrock}-identified QSOs due to an increased contribution from host-galaxy emission and/or dust extinction. The high recovery rate of the missed QSOs from the modified pipeline provides a larger QSO yield by identifying both lower luminosity QSOs (i.e.,\ analogous to distant ``Seyfert galaxies") and more dust-reddened QSOs than the standard pipeline. See \S5.1.
    \item A diverse range of QSOs are selected within the DESI QSO survey including host-galaxy dominated QSOs, dust-reddened QSOs, BALQSOs, narrow-line QSOs, and QSOs with intervening absorption features. In addition to providing important cosmological tracers, the DESI quasar survey reveals a large and diverse sample of QSOs for a broad range of astrophysical studies. See \S5.2.
    \item The vast majority of the galaxy contaminants do not appear to host QSO or AGN activity. Their strong UV spectral slopes and prominent host-galaxy signatures suggest they may be lower-luminosity analogs to compact post-starburst galaxies identified in the SDSS. See \S5.3.
\end{itemize}

The presented VI dataset is a high-quality resource for quantifying the key metrics of the DESI survey to ensure that all users of the DESI data can be confident in the results and fully understand the data quality and any potential issues. Following each data-assembly release we will re-visually inspect the spectra around the low-quality--high-quality threshold (VI~$\approx$~2.5) to further improve the quality of the overall VI dataset and, consequently, use these refined data to re-compute the key survey metrics. Several other VI efforts are also ongoing within the DESI survey, focused on assessing the redshift and spectral classification quality for specific scientific projects, in addition to further testing and validation of our ``afterburner" identification approaches (following our sparse VI; see Table~\ref{table:sparse_subsets}) with the objective of further improving the performance of the DESI quasar survey.

\section*{Data Availability}
All data points shown in the figures are available in machine-readable form from https://doi.org/10.5281/zenodo.7316969

%\begin{acknowledgements}
%\section*{Acknowledgment}
\vspace{\baselineskip}
We thank the anonymous referee for their thoughtful comments. DMA and VAF acknowledge the Science Technology and Facilities Council (STFC) for support through grant codes ST/T000244/1 and ST/S505365/1 (quota studentship). TWL acknowledges support from the Ministry of Science and Technology (MOST 111-2112-M-002-015-MY3), the Ministry of Education, Taiwan (Yushan Young Scholar grant NTU-110VV007), National Taiwan University research grant (NTU-CC-111L894806). ADM acknowledges support by the U.S. Department of Energy, Office of Science, Office of High Energy Physics, under Award Number DE-SC0019022.

This research is supported by the Director, Office of Science, Office of High Energy Physics of the U.S. Department of Energy under Contract No. DE–AC02–05CH11231, and by the National Energy Research Scientific Computing Center, a DOE Office of Science User Facility under the same contract; additional support for DESI is provided by the U.S. National Science Foundation, Division of Astronomical Sciences under Contract No. AST-0950945 to the NSF’s National Optical-Infrared Astronomy Research Laboratory; the Science and Technologies Facilities Council of the United Kingdom; the Gordon and Betty Moore Foundation; the Heising-Simons Foundation; the French Alternative Energies and Atomic Energy Commission (CEA); the National Council of Science and Technology of Mexico (CONACYT); the Ministry of Science and Innovation of Spain (MICINN), and by the DESI Member Institutions: https://www.desi.lbl.gov/collaborating-institutions.

The DESI Legacy Imaging Surveys consist of three individual and complementary projects: the Dark Energy Camera Legacy Survey (DECaLS), the Beijing-Arizona Sky Survey (BASS), and the Mayall z-band Legacy Survey (MzLS). DECaLS, BASS and MzLS together include data obtained, respectively, at the Blanco telescope, Cerro Tololo Inter-American Observatory, NSF’s NOIRLab; the Bok telescope, Steward Observatory, University of Arizona; and the Mayall telescope, Kitt Peak National Observatory, NOIRLab. NOIRLab is operated by the Association of Universities for Research in Astronomy (AURA) under a cooperative agreement with the National Science Foundation. Pipeline processing and analyses of the data were supported by NOIRLab and the Lawrence Berkeley National Laboratory. Legacy Surveys also uses data products from the Near-Earth Object Wide-field Infrared Survey Explorer (NEOWISE), a project of the Jet Propulsion Laboratory/California Institute of Technology, funded by the National Aeronautics and Space Administration. Legacy Surveys was supported by: the Director, Office of Science, Office of High Energy Physics of the U.S. Department of Energy; the National Energy Research Scientific Computing Center, a DOE Office of Science User Facility; the U.S. National Science Foundation, Division of Astronomical Sciences; the National Astronomical Observatories of China, the Chinese Academy of Sciences and the Chinese National Natural Science Foundation. LBNL is managed by the Regents of the University of California under contract to the U.S. Department of Energy. The complete acknowledgments can be found at https://www.legacysurvey.org/.

The authors are honored to be permitted to conduct scientific research on Iolkam Du’ag (Kitt Peak), a mountain with particular significance to the Tohono O’odham Nation.
%\end{acknowledgements}

{}

\end{document}